\documentclass[fleqn,usenatbib]{mnras}

\usepackage{newtxtext,newtxmath}
\usepackage[T1]{fontenc}

\DeclareRobustCommand{\VAN}[3]{#2}
\let\VANthebibliography\thebibliography
\def\thebibliography{\DeclareRobustCommand{\VAN}[3]{##3}\VANthebibliography}

\usepackage{graphicx}	% Including figure files
\usepackage{amsmath}	% Advanced maths commands
\title[Slingshot prominences]{Slingshot prominences: a hidden mass loss mechanism}

\author[R. F. P. Waugh et al.]{
Rose F.P. Waugh,$^{1}$\thanks{E-mail: rw47@st-andrews.ac.uk (RFPW)}
Moira M. Jardine,$^{1}$
J. Morin,$^{2}$, and
J.-F. Donati$^{3}$\\
$^{1}$School of Physics and Astronomy, University of St Andrews, North Haugh, St Andrews, Fife, Scotland, KY16 9SS\\
$^{2}$LUPM, Universit\'e de Montpellier, CNRS, Place Eug\`ene Bataillon, F-34095 Montpellier, France\\
$^{3}$IRAP, Universit\'{e} de Toulouse, CNRS, UPS, CNES, 14 Avenue Edouard Belin, 31400, Toulouse, France\\
}

\date{Accepted XXX. Received YYY; in original form ZZZ}

\pubyear{2020}

\begin{document}
\label{firstpage}
\pagerange{\pageref{firstpage}--\pageref{lastpage}}
\maketitle

\begin{abstract}
% MAX 250 words
Whilst ``slingshot'' prominences have been observed on M-dwarfs, most if not all theoretical studies have focused on solar-like stars. We present an investigation into stellar prominences around rapidly rotating young M-dwarfs. We have extrapolated the magnetic field in the corona from Zeeman-Doppler maps and determined the sites of mechanical stability where prominences may form. We analyse the prominence mass that could be supported and the latitude range over which this material is distributed. We find that for these maps, much of this prominence mass may be invisible to observation - typically <1\%  transits the stellar disc. On the rapidly-rotating M-dwarf V374 Peg (P$_{\rm rot}$ = 0.45 days) where prominences have been observed, we find the visible prominence mass to be around only 10\% of the total mass supported.  The mass loss rate per unit area for prominences scales with the X-ray surface flux as $\dot{M}/A \propto$ $F_X^{1.32}$ which is very close to the observationally-derived value for stellar winds. This suggests that prominence ejection may contribute significantly to the overall stellar wind loss and spin down. A planet in an equatorial orbit in the habitable zone of these stars may experience intermittent enhancements of the stellar wind due to prominence ejections. On some stars, this may occur throughout 20\% of the orbit. 
\end{abstract}


%%%%%%%%%%%%%%%%%%%%%%%%%%%%%%%%%%%%%%%%%%%%%%%%%%

%%%%%%%%%%%%%%%%% BODY OF PAPER %%%%%%%%%%%%%%%%%%

\section{Introduction}
Stellar prominences are condensations of coronal plasma supported by the stellar magnetic field. Unlike prominences on the Sun, ``slingshot prominences'', which are found on rapidly rotating stars, are significantly larger (10-100 times the mass~\citep{Cameron1999}) and co-rotate with the star at greater distances  - typically a few stellar radii above the surface. These prominences have been observed over many years on the young star AB Doradus [$M_*=0.86M_{\odot}$~\citep{Innis1988}, $P=0.514$days~\citep{Guirado2010}], where they were first detected~\citep{CC1989}. The clouds are observed in the H$_{\alpha}$ line profile as absorption dips that cross the stellar disc in a matter of hours, and sometimes reappear on consecutive nights. ~\cite{CC1989} deduced that these features originate from material co-rotating with the star, close to or beyond the stellar co-rotation radius. They explained these features as the presence of large condensations of hydrogen, supported by the strong magnetic fields found on such rapidly rotating, young stars. The existence of these co-rotating condensations requires that the stellar magnetic field must be closed in these locations, in order to support the material.  Since then, slingshot prominences have been found on many more rapidly rotating stars~\citep{CCWoods92,Jeffries93,ByrneEibe1996,Barnes98,Eibe98,Barnes2000,Dunstone2005,DunstoneII,Skelly2008,Skelly2009,Skelly2010,Leitzinger2016}. These range from other solar-like stars, to T Tauri stars and M-dwarfs. The  variety and number of stars on which these features have been observed implies that these prominences could be common amongst young stars.\\ 

Line shifts that could indicate the destabilisation of prominence material have also been observed \citep{2014MNRAS.443..898L,2017IAUS..328..198K,2019A&A...623A..49V}. Slingshot  prominences are typically supported at or beyond the co-rotation radius and so will be centrifugally ejected if they lose magnetic support. Solar-like prominences, in contrast, that are destabilised below the co-rotation radius, may fall passively back towards the surface, or they may also be ejected. Line asymmetries alone cannot distinguish between these two cases. 
In a large survey of such line asymmetries, however, \citet{2019A&A...623A..49V} comment that most of the observed velocites were below the escape speed and so would not be expected to contribute to the overall mass loss from the star.\\

The ultrafast rotator and M-dwarf V374 Peg has also been observed to host prominences~\citep{Vida16} and potentially related features were observed in the K2 dips of 19 other M-dwarfs, in a study by~\cite{Stauffer17}. K2 data of these 19 stars showed absorption dips in the light curve that repeated with the stellar rotation period, suggesting the features were co-rotating with the stars. M-dwarfs are the most numerous spectral type within the Milky Way, comprising of about 70\% of stars~\citep{Bochanski2010}. They are small enough to be fully convective, and typically show strong, simple stellar magnetic fields which are ideal for supporting prominences. At the very lowest masses, however, some M-dwarfs may also exhibit weak but complex field structures, suggesting that their dynamos may exist in a ``bistable'' regime~\citep{Morin11}. The fact that M-dwarfs can be classified into these two groups based on their field structure makes them excellent tests of models of prominence support, as results can then be compared within the same spectral type. 

Due to their cooler temperatures and lower luminosity than more massive stars, the habitability zone of M-dwarfs is closer in to the star. Observations suggest that rocky exoplanets in orbit around M-dwarfs are in close orbits, with about a third of these found to lie within the habitability zone~\citep{shields2017} where they are capable of supporting liquid water. The extended lifetimes of M-dwarfs, due to the mixing that can occur in fully convective stars allowing a larger supply of hydrogen for burning, also contributes to them being good candidates for hosting life-bearing worlds. However,~\cite{Khodachenko} theorised that exposure to CME (coronal mass ejection) material for exoplanets within an M-dwarf habitable zone was a serious and continuous threat, throughout the entire life of the star, and this does not add to the list of positive conditions for habitability. Prominences are known to be ejected from the Sun if they become unstable and this is not specific to our star (~\citealt{Haisch83},~\citealt{Hussain2013}, and references within~\citealt{ACStellProm}) and these could also be ejected into the path of orbiting planets. Whilst CMEs and prominences are distinct features, ejections of large solar prominences are often observed to accompany CMEs. The  masses of stellar slingshot prominences may therefore be used to give a lower limit to stellar CME mass. It is the highly energetic particles produced by CMEs that are able to strip atmospheres~\citep{Mars2015} and prominences, which do not contain these, are unlikely to do as much damage. However, on stars that are predicted to produce regular prominence ejections ~\citep{WindGauges}, large quantities of ejected prominence material could frequently bombard the planet. The consequences of this are not well understood.\\

Both of these forms of stellar ejecta are related to the stellar magnetic field and although magnetic fields can only be detected at the surface of the star, previous studies have shown that slingshot prominences could be useful in testing the methods used to construct the coronal magnetic field structure of stars~\cite{Jardine19}. These methods rely on extrapolation of the surface magnetic field vectors that are obtained from Zeeman Doppler Imaging (ZDI). The coronal field can be constructed from this surface field by assuming the field to be of a particular form, such as force free or potential. 
~\cite{Jardine19} showed, through modelling the prominence locations and synthesising H$_{\alpha}$ spectra, that a potential field reproduced the observed absorption features well, unlike the non-potential field. 
%Thus, the coronal field topology that was produced by assuming the field to be potential must have matched the reality better than the non-potential model. 
Typically, in addition to the map of the radial magnetic field at the surface, an upper boundary condition is also required in order to define the field structure. This upper boundary condition can be taken as the ``source surface'', the radius at which the magnetic field becomes purely radial. Work in determining the location of this radius for stars other than the Sun was published by~\cite{Reville15}, where stellar wind models were used to determine the opening of field lines by the wind, and thus the location of the source surface. The location of the source surface was also considered by~\cite{See18}, who looked at the open flux determined from the ZDI maps and a potential field extrapolation, in order to model the evolution of angular momentum of solar like stars. Using the accepted solar source surface of $2.5R_{\odot}$, they then modelled the relationship between the source surface and stellar rotation period.  Prominences may also be useful in tracing regions of locally-closed magnetic field, since the support of a prominence requires that the magnetic field must be closed at that point. At other places around the star, however, the magnetic field may be open at this radius. Studies have shown that prominences are able to form within such open regions by magnetic reconnection of field lines if there is equilibrium available for the magnetic loop~\citep{JardineCC91,JardineB05,Waugh18}.\\

%The large masses of these slingshot prominences, combined with evidence from AB Dor that they are likely to be numerous on a each star [REF], suggests that prominence ejection on these stars could be an important mass loss mechanism [REF].
The dominant form of mass and angular momentum loss for a star is usually taken to be the stellar wind~\citep{WeberDavis67,Vidotto11}. This loss of angular momentum will influence the spin down rate of the star, with the spin down rate and magnetic field being linked through the action of the stellar dynamo. Thus, the evolution of the star on the main sequence is governed by the angular momentum and mass loss mechanisms. Large prominences around young, rapid rotators could also be mechanisms for sizeable mass and angular momentum loss if ejected from the star~\citep{CC1989,VdA18,VdA19,Jardine19}. ~\cite{VdA19} showed that for a solar like star these prominences reach their maximum mass and lifetime once the star reaches the zero-age main-sequence but that on fast rotators these prominences could be supported up to an age of 800Myrs. During this time they will be contributing to the mass and angular momentum loss to varying degrees.\\

Recently,~\cite{WindGauges} presented a novel method of predicting stellar wind mass loss rates by using the prominences. The winds of cool stars are notoriously difficult to measure but are crucial to our understanding of the star's evolution. The model uses the slingshot prominences as ``wind gauges''. Prominences are formed by the up-flow of hot plasma along flux tubes. This plasma then condenses at stable points well above the stellar surface. The magnetic field lines that support these prominences act as nets to catch the up-flow of material that is supplied by this isothermal flow, very similar to the stellar wind. Since the masses and lifetimes of these prominences can be observed, this provides a method of predicting a mass loss rate which is likely to be similar to the wind regions of the corona. Their paper assumes the area of the surface contributing to the prominences to be 1\%, though the extent of the stellar surface that is feeding a prominence may vary from star to star.\\

%Eruption of solar prominences and coronal mass ejections (CMEs) are solar events with consequences for Earth. With slingshot prominences at 10-100 times the mass as solar ones, and in co-rotation with the star at multiple stellar radii from the stellar surface, the consequences of a prominence ejection around a young, rapidly rotating star could be more damaging to an orbiting planet than would be experienced on Earth [REF]. \\

The studies conducted so far that model these features focus on solar-like stars~\citep{Ferreira2000,JardineB05,VdA18,Waugh18,VdA19}. This has been partly driven by the need to understand the role these prominences played in our Sun's past, but also because the prototype star and candidate for which there is the most observational data to compare to, AB Dor, is also solar-like. Here we investigate the formation sites, visibility and consequences of prominences on a selection of M-dwarfs for which we have access to the Zeeman Doppler Images (ZDI), or surface magnetic field maps. For young stars, where prominences are more massive than their solar counterparts, the accumulative mass loss rate from regular ejection of the supported prominences may not be negligible and could have consequences for the stellar evolution. 

%\newpage
\section{Method}
Zeeman Doppler Imaging maps of a set of M-dwarfs are used to reconstruct the stellar magnetic field, assuming it to be potential. All of the sites of stable mechanical equilibrium within this field are determined. These stable points represent possible prominence formation sites for these maps. Once the formation sites for these maps are found, their properties (such as their mass and visibility) are determined. We refer to these as ``prominence formation sites'' or ``predicted prominences'' throughout this paper. This is to emphasise that the features in this paper are predictions, based on the observed ZDI maps which are used as an input in constructing the coronal magnetic field. \\

The observability of prominences is governed by the nature of the H$\alpha$ source function, which for prominences is almost pure scattering. As a result, when prominences transit the stellar disc, they scatter H$\alpha$ photons out of the line of sight, producing characteristic absorption transients that move through the H$\alpha$ line profile. When prominences  are out of transit, they {\it can} be detected in emission. In this case, however, the geometrical dilution of the stellar flux is such that this emission is very hard to detect, unless these prominences are very close to the stellar surface~\citep{Odert2020}. Given that most prominences form at or around the co-rotation radius, this means that these stars must be very rapid rotators. There are a  few notable examples where prominence emission features are seen  (such as LQ Lup~\citep{Eibe98} and Speedy Mic~\citep{DunstoneII}). In the case of stars such as LQ Lup, where the inclination of the stellar rotation axis is so large that prominences never pass behind the star, all of the prominences can be seen, allowing a complete census of the total prominence mass. The cases where prominences are seen in emission are in the minority, however, and so for the rest of this paper we consider as  ``visible'' or ``observable'' only those prominences that transit the star.\\

\subsection{The stellar sample}
The stars in our stellar sample are all M-dwarfs and their masses, radii, periods and inclinations are given in Table~\ref{tab:stellardata}. For some of these stars, maps for multiple years have been used, whilst for others there were only maps from one year available to us. This sample represents a subset of a larger survey of the magnetic fields of M-dwarfs ~\citep{Morin10,Morin08,Donati08} where we have selected those stars with the smallest co-rotation radii, since these are the ones most likely to support slingshot prominences.

%--------TABLE--------
\begin{table}
\begin{center}
\begin{tabular}{ccccc}
    \hline Maps & $M$ [$M_{\odot}$] & $R$ [$R_{\odot}$] & $P$ [days] & $i$ [deg]\\ \hline
    EQ Peg B (2006) & 0.25 & 0.25 & 0.40 & 60\\
    GJ1156 (2007/08/09) & 0.14 & 0.16 & 0.33 & 40\\
    AD Leo (2007/08) & 0.42 & 0.38 & 2.24 & 20 \\
    EQ Peg A (2006) & 0.39 & 0.35 & 1.06 & 60\\
    GJ1111 (2007/08) & 0.10 & 0.11 & 0.46 & 60\\
    GJ1245b (2006/07/08) & 0.12 & 0.14 & 0.71 & 40\\
    GJ9520 (2008) & 0.55 & 0.49 & 3.40 & 45\\
    GJ182 (2007) & 0.75 & 0.82 & 4.35 & 60\\
    GJ494 (2007/08) & 0.59 & 0.53 & 2.85 & 60\\
    V374 Peg (2006) & 0.30 & 0.35 & 0.45 & 70\\
    \hline
\end{tabular}
\caption{Table of parameters for the stars used here. Symbols here are for the stellar parameters; $M$ for mass, $R$ for radius, $P$ for period and $i$ for inclination~\citep{Morin10,Morin08,Donati08}.}
\label{tab:stellardata}
\end{center}
\end{table}
%----------------------

\subsection{ZDI maps predict prominence formation sites}
The ZDI maps provide the magnetic field strength and direction at the stellar surface. We assume the field to be potential, i.e. $\nabla\times\underline{B}=\underline{0}$ and since $\nabla.\underline{B}=0$ this yields Laplace's equation $\nabla^2\Psi=\underline{0}$, where $\Psi$ is the flux function. This can be solved in spherical polar coordinates (R,$\theta$,$\phi$) by separation of variables to give a solution of spherical harmonics:

\begin{equation}
    \Psi = \sum_{l=1}^N\sum_{m=-l}^l [a_{lm}R^l + b_{lm}R^{-(l+1)}]P_{lm}(\theta)e^{im\phi}
\end{equation}
where $P_{lm}$ are the associated Legrendre polynomials. The magnetic components are then given by;
\begin{equation}
    B_R = -\sum_{l=1}^N\sum_{m=-l}^l [la_{lm}R^{l-1} -(l+1)b_{lm}R^{-(l+2)}]P_{lm}(\theta)e^{im\phi}
\end{equation}
\begin{equation}
    B_{\theta} = -\sum_{l=1}^N\sum_{m=-l}^l [a_{lm}R^{l-1} + b_{lm}R^{-(l+2)}]\frac{d}{d\theta}\Bigr(P_{lm}(\theta)e^{im\phi}\Bigr),
\end{equation}
\begin{equation}
    B_{\phi} = -\sum_{l=1}^N\sum_{m=-l}^l [a_{lm}r^{l-1}+ b_{lm}R^{-(l+2)}]\frac{P_{lm}(\theta)}{\sin{\theta}}ime^{im\phi}.
\end{equation}
$a_{lm}$ and $b_{lm}$ are coefficients, determined by the boundary conditions. The radial field at the surface is known, and another boundary condition can be constructed by assuming that at some height, $R_{ss}$, the field becomes open and radial i.e. $B_{\theta}(R_{ss})=B_{\phi}(R_{ss})=0$. This is the ``source surface''. In this work, this is set at $R_{ss}=18R_*$
for all stars. This was chosen to ensure that the field is closed at the co-rotation radius for all stars in our sample and ensures the possibility of forming condensations on each map. Prominences are typically found around the co-rotation radius of stars such as AB Dor and Speedy Mic, therefore in this modelling we are interested especially in any prominence mass that may be supported around the stars' co-rotation radii. In order to support any prominence material in this modelling, the magnetic field must be closed at this point. Some stars in this sample have a large co-rotation radius and we chose to set the source surface in this model at a location that would encompass the co-rotation radii of all stars in the sample so that it would be consistent across our sample.
The field is then constructed using code developed initially by van Ballegooijen, Cartledge $\&$ Priest in 1998 for studying filament formation on the Sun~\citep{Ballegooijen98}.\\

%--------FIGURE--------
\begin{figure}
    \centering
    \includegraphics[width=\columnwidth]{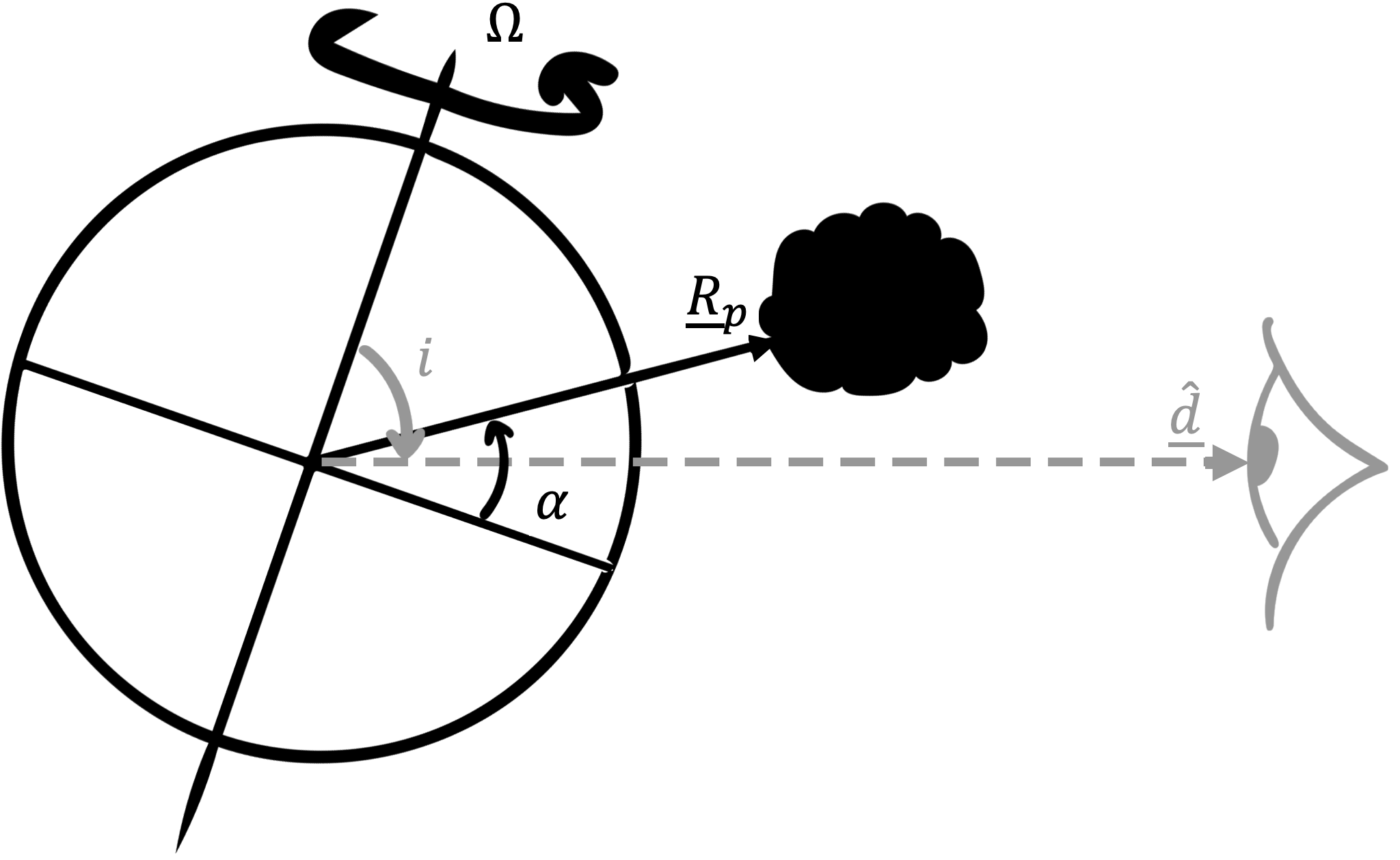}
    \caption{Cartoon of the set-up for checking the visibility.}
    \label{fig:anglecartoon}
\end{figure}
%---------------------

The equilibrium points are points in which the forces acting on the magnetic loop are equal, i.e. the gas pressure variation, magnetic, gravitational and centrifugal forces are balanced;
\begin{equation}
    \underline{0} =  - \nabla{p} + (\textbf{\textit{B}}.\nabla)\frac{\textbf{\textit{B}}}{\mu} - \nabla\Bigr(\frac{B^2}{2\mu}\Bigr) + \rho\textbf{\textit{g}},
    \label{eqn:eom}
\end{equation}
where $p$ is the gas pressure, $\textbf{\textit{B}}$ is the magnetic field, $\mu$ is the permeability of free space, $\rho$ is the gas density and $\textbf{\textit{g}}$ is the effective gravity. The effective gravity is the combination of the gravitational and centrifugal forces,
\begin{equation}
    \textbf{\textit{g}} = \Bigr(-\frac{GM_{\star}}{R^2}+\Omega^2R\sin{\theta}\Bigr)\hat{R} + \Bigr(\Omega^2R\sin{\theta}\cos{\theta}\Bigr)\hat{\theta},
\end{equation}
and the co-rotation radius is the distance from the stellar centre, typically within the equatorial plane, at which this is zero. Here the gravitational and centrifugal forces balance.\\

\cite{Ferreira2000} showed that to satisfy mechanical equilibrium,  $\underline{g}.\underline{B}=0$. Possible prominence formation sites are those equilibrium points that are mechanically stable. The stability of these requires that the component of the effective gravity along the magnetic field line is decreasing, i.e. we find a potential minimum. Mathematically this is written as
\begin{equation}
    (\textbf{\textit{B}}.\nabla)(\textbf{\textit{g}}.\textbf{\textit{B}})<0.
\end{equation}

A prominence is assumed to form at each stable point and to have the maximum possible density for support, given by $\rho g = B^2/\mu R_c$ where $R_c$ is the local radius of curvature of the field~\citep{VdA18}. Prominence lifetimes are determined from the time taken for a thermal wind to supply this mass. The temperature of the corona is set at $8.57\times10^6$K~\cite{Cameron1999} for all stars.

%--------FIGURE--------
\begin{figure*}
    \centering
    \includegraphics[width=\textwidth]{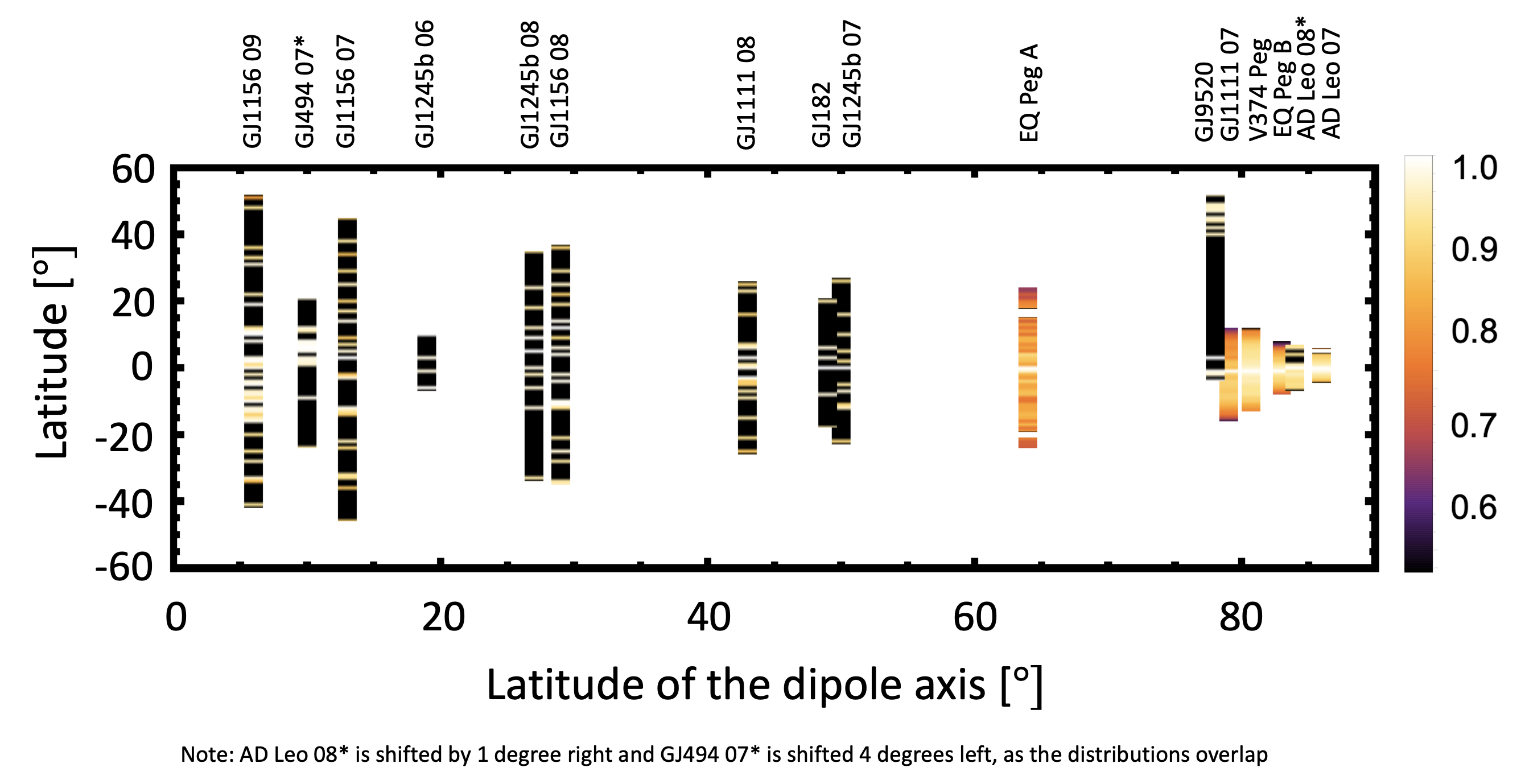}
    \caption{The distribution of latitudes of prominence formation sites that are supported for each map. The colour table shows the mass distribution of prominence material over these latitudes, scaled to the maximum value supported on each star. The maximum values are shown in Table~\ref{tab:vismassdata}.}
    \label{fig:massdistribution}
\end{figure*}
%---------------------

\subsection{Checking the visibility of formation sites}
\label{sec:vis}
The visibility of a prominence around a star from our vantage point on Earth is dependent on the inclination of the star's rotation axis. The visible locations are those that transit the star.
We consider (in Cartesian coordinates) the instantaneous prominence formation site, $\textbf{\textit{R}}_p=(x_p, y_p, z_p)$, which by coordinate transformation becomes:
\begin{equation}
    \textbf{\textit{R}}_p = |\textbf{\textit{R}}_p|(\cos(\lambda)\cos(\alpha), \sin(\lambda)\cos(\alpha), \sin(\alpha)),
\end{equation}
where $\lambda$ is the phase of the prominence at time t = 0 and $\alpha$ is its latitude (see Appendix A). The line of sight vector is defined by
\begin{equation}
\hat{\textit{\textbf{d}}} = (d_x,  d_y,  d_z)
\end{equation}
which, in terms of the stellar inclination ($i$) and rotation phase ($\Omega t$) is
\begin{equation}
   \hat{\textit{\textbf{d}}} = (\cos(-\Omega t)\sin(i), \sin(-\Omega t)\sin(i), \cos(i)).
\end{equation}
The locations at which a prominence would cross the stellar disc can be found by considering the distance cosine, 
\begin{equation}
    \cos(\phi) = \frac{\textit{\textbf{R}}_p.\underline{\hat{\textit{\textbf{d}}}}}{|\textit{\textbf{R}}_p|},
 \end{equation}
which can be written as
\begin{multline}
    \cos(\phi) = \cos(\lambda)\cos(\alpha)\cos(\Omega t)\sin(i)\\ - \sin(\lambda)\cos(\alpha)\sin(\Omega t)\sin(i) + \sin(\alpha)\cos(i).
\end{multline}
We require
\begin{equation}
R_* = |\textbf{\textit{R}}_p|\sin(\phi)=R_p\sqrt{1-\cos^2(\phi)}
\end{equation}
i.e. \(\cos(\phi) = \sqrt{1-(R_*/R_p)^2}\) which can be written as:
\begin{equation}
    \cos(\Omega t+\lambda) = \frac{\sqrt{1-(R_*/R_p)^2}- \sin(\alpha)\cos(i)}{\cos(\alpha)\sin(i)}.
    \label{eqn:phases}
\end{equation}
This must be $\leq$ 1 for prominence visibility. We solve this for latitude ($\alpha$) in terms of distance from the rotation axis ($R_p$). On visibility plots throughout this paper, the visible regions are shown in white and the locations that would not be visible are shaded out in grey. The locations of the predicted prominences from our modelling are then superimposed on top of these visibility plots.

\subsection{Mass loss and angular momentum loss rates}

The mass loss rate associated with the prominences can be found by considering the flow of material along these closed prominence bearing loops, since this is the mechanism by which the prominence formation sites fill up. The stars in our sample lie in a \textit{limit-cycle regime} \citep{WindGauges} since their co-rotation radii (where prominences form) lie beyond the sonic point for up-flows from the surface. As a result, the surface will continually supply mass, even once the prominence reaches the maximum mass that can be supported and is ejected. Prominences will repeatedly form and be ejected and their time-averaged mass loss rate will be equal to the mass flow rate from the surface. This is given by
\begin{equation}
    \dot{M}_\mathrm{prom}, =\rho_p u_p A_p
\end{equation}
where $\rho_p$ is the prominence density, $u_p$ the up-flow velocity into the prominence and $A_p$ the area contributing to the prominence. By mass conservation, this can be calculated at any radius from the star. Here we evaluate it at the stellar surface. The up-flow that supplies the prominences is taken to be isothermal, with the coronal temperature of $T = 8.57\times10^6K$. The density is calculated from the plasma pressure, which is estimated by the scaling of the surface gas and magnetic pressures $p= \kappa B^2$, where $\kappa = 10^{-5}$~\citep{Jardine19}.\\

The prominences themselves are set at a temperature of $T_p=8500K$, the temperature suggested by~\cite{CC1990} for prominences on AB Dor. The masses of the prominences can be found by summing the mass at each point along the field line, which is dependent on the local density.\\

The angular momentum loss rate, $\dot{J}_\mathrm{prom}$, can also be estimated:
\begin{equation}
    \dot{J}_\mathrm{prom} = \Omega (R_p \sin{\theta})^2\dot{M}_\mathrm{prom}. 
\end{equation}
We note that this neglects the magnetic stresses acting on the prominence as it is ejected and so represents a lower limit on the angular momentum loss rate.

%\subsection{Mass loss rate as seen by an orbiting planet}
%Finally, we consider how this prominence mass loss rate could be experienced by an orbiting planet. We consider an example Earth-sized planet %(R$_{\oplus}$) 
%\textcolor{red}{orbiting at 0.05AU from its star (this is the predicted habitable zone the Earth-mass planet Proxima b  - see 2021 paper by JF's student Baptist Klein for details)}. This planet subtends \textcolor{red}{estimate needs done more carefully - 0.1$^\circ$ as seen from the star.}
%This orbits at a distance of less than 0.05AU from its star}at a distance of 1AU from the star, as is depicted in the cartoon in Figure~\ref{fig:starplanet}. The cone subtended by the planet at this orbit is range of latitudes from which prominence material on the star could impact on the planet is $\pm x^o$. We estimate this using simple trigonometry as:
%\begin{equation}
%    x \approx \arctan{R_{\oplus}/R_{\star}}
%    = 0.5^o.
%\end{equation}
%From this, we calculate the total mass loss rate interacting with the planet and the percentage of the orbit that would be affected.
%--------FIGURE--------
%\begin{figure}
%    \centering
%    \includegraphics[width=\columnwidth]{starplanet.png}
%    \caption{Cartoon showing an estimate of the latitude, x, over which the prominence material could interact with an orbiting %planet.\textcolor{red}{I think this figure needs a revamp}}
%    \label{fig:starplanet}
%\end{figure}
%---------------------

%\newpage
\section{Hosting observable prominences}
%\newpage
\subsection{Prominence formation sites depend on the tilt of the dipole axis}
The distribution of latitudes at which prominence formation sites are supported on each star is shown in Figure~\ref{fig:massdistribution}. The colour table on this plot also shows the mass distribution over these latitudes. There is a trend for maps that show their dipole axis to be at lower latitudes (i.e. those with dipole axes that are more tilted relative to the rotation axis) to support stable points over a larger range of latitudes than those whose dipole axes are aligned with the rotation axis. Whilst these tilted fields are able to support high latitude prominences, the mass supported at these latitudes is small. The maps on the left hand side of this plot show smooth distributions in mass over the small latitude range where they could support prominence material, whilst those on the right hand side with tilted dipoles show more clumpy mass distributions. The largest masses are generally supported close to the equatorial plane, where the centrifugal term in the effective gravity is largest 
%Since the centrifugal force is largest in the equatorial plane, the pressure gradient is steepest here which results in larger prominence masses
~\citep{JardineCC91,JardineB05,Waugh18}.  \\

The stars on the left hand side of Figure~\ref{fig:massdistribution} are those that lie in the \textit{bistable} regime. These are very low mass stars which show weak and complex magnetic fields. The combination of these factors enables them to support less prominence mass than other stars. These factors make it difficult to find stable points in the field, and result in lower masses and a more ``clumpy'' distribution of mass than is seen on the right hand side of the plot.

%\newpage
\subsection{Visibility of prominence formation sites}

Whilst every map investigated here has been shown to support stable points (prominence formation sites), the vast majority of these would likely not be visible due to the geometry of the system. In order for a prominence to be visible to us observing from Earth, it must cross the stellar disc, as discussed in section~\ref{sec:vis}. The visibility of a prominence will clearly depend on where the prominence forms and which areas around the star will be visible to us. Figure~\ref{fig:VisEQPegA} shows an example ``visibility plot'' for the map of EQ Peg A. The grey regions are the positions in space around the star which will not cross the stellar disc. The white region in the centre shows the locations around a star which would cross the stellar disc. The optimal latitude for prominence formation would be $(90-i)$, as this latitude will cross across the centre of the the stellar disc and thus be visible regardless of the radius at which the prominence forms. For all other latitudes, there will be a radius at which the prominence would no longer cross the stellar disc. In this visibility plot, the prominences found are shown alongside the co-rotation radius of the star (the purple dashed line). The colour scale represents the mass of the prominences, scaled to the maximum prominence mass supported. The largest prominences can be seen to form around the equatorial co-rotation radius - which is reflected in all maps. Appendix~\ref{app:vis} shows the visibility plots for the remaining maps. Beyond the equatorial co-rotation radius, the mass of the prominences decreases extremely rapidly. As the prominences typically form at or beyond the co-rotation radius and the co-rotation radii for these stars are reasonably large, the prominences on these stars would mostly not be visible to us. Good stellar candidates for supporting visible stellar prominences would have the following features:

%--------FIGURE--------
\begin{figure}
    \centering
    \includegraphics[width=1\columnwidth]{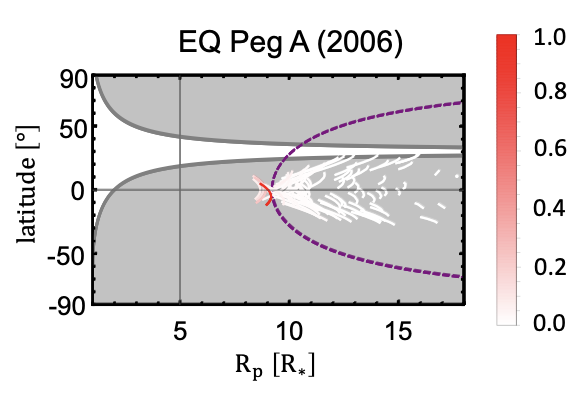}
    \caption{Visibility plot for the map of EQ Peg A (2006). The grey regions are those that would not cross the stellar disc - i.e. not be visible. The purple dashed line shows the co-rotation radius and the prominences are shown coloured by mass, with 1 being the maximum prominence mass supported on the star.}
    \label{fig:VisEQPegA}
\end{figure}
%---------------------

%------TABLE--------
\begin{table}
\begin{center}
 \begin{tabular}{|c c c c|}
 \hline
 Map & Max prom. mass in& Total prom. mass & \% mass \\
 & $1^\circ$ latitude band [kg] & supported [kg] & visible \\
 \hline
 EQ Peg B 2006 & 2.6$\times10^{16}$ & 3.7$\times10^{16}$ & 0 \\ 
 GJ1156 2007 & 1.2$\times10^{12}$  & 2.9$\times10^{12}$ & 0.1 \\
 GJ1156 2008 & 7.4$\times10^{11}$ & 3.5$\times10^{12}$ & 0 \\
 GJ1156 2009 & 5.9$\times10^{11}$ & 4.8$\times10^{12}$ & 0.2 \\
 AD Leo 2007 & 1.3$\times10^{14}$  & 2.3$\times10^{14}$ & 0 \\
 AD Leo 2008 & 2.4$\times10^{14}$  & 2.6$\times10^{14}$ & 0 \\
 EQ Peg A 2006 & 1.9$\times10^{15}$  & 3.6$\times10^{15}$ & 0.1  \\
 GJ1111 2007 & 2.0$\times10^{13}$  & 2.6$\times10^{13}$ & 0 \\
 GJ1111 2008 & 1.3$\times10^{11}$ & 3.4$\times10^{11}$ & 0.7 \\
 GJ1245b 2006 & 5.2$\times10^{11}$  & 1.2$\times10^{12}$ & 0 \\
 GJ1245b 2007 & 2.0$\times10^{12}$  & 2.9$\times10^{12}$ & 0 \\
 GJ1245b 2008 & 9.9$\times10^{9}$  & 3.1$\times10^{10}$ & 0 \\
 GJ9520 2008 & 4.3$\times10^{12}$  & 1.4$\times10^{13}$ & 22.9 \\
 GJ182 2007 & 5.1$\times10^{13}$ & 8.7$\times10^{13}$ & 0 \\
 GJ494 2007 &  8.3$\times10^{12}$ & 3.5$\times10^{13}$ & 50.4\\
 GJ494 2008 & 6.8$\times10^{8}$&  6.8$\times10^{8}$ & 0 \\
 V374 Peg 2006 & 6.5$\times10^{17}$  & 5.1$\times10^{17}$ & 12.9 \\
 \hline
\end{tabular}
 \caption{Table showing the maximum prominence mass supported in any $1^o$ latitude band (i.e. 1 in the colour band in Figure~\ref{fig:massdistribution}, the total prominence mass supported in each star and the percentage of this that is visible.}
\label{tab:vismassdata}
\end{center}
\end{table}
%-----------------

\begin{itemize}
    \item They would have small co-rotation radii to allow for prominence formation at low heights above the stellar surface. This increases the range of prominence latitudes that would be visible. Observationally, stars with high equatorial velocities would be good targets as this will result in low co-rotation radii.
    \item They would have high stellar inclinations such that we observe the star close into the equatorial plane, where most prominences form. This is also typically where the most massive prominences would form, which would be the easiest to see in the observational H$_{\alpha}$ spectra as these prominences will scatter the most light.
    \item Cases where the dipole axis has a low latitude which supports high latitude prominences could also be good candidates. The stellar inclination is a fixed parameter when observing a star, but if the star supports high latitude prominences then these may still be visible even if the stellar inclination is low.
\end{itemize}

We do not expect that the latitude of the dipole axis will be constant and indeed this is borne out in the cases where there are maps obtained over consecutive years. The field structure of the Sun is known to vary cyclically over a 22 year period, and it seems likely that other solar-like stars show similar cyclical behaviour. The timescales for such stellar cycles are likely not the same as the solar cycle; Jeffers recently investigated the stellar cycle of the star Tau Boo and found a period of 120 days~\citep{Jeffers18}, whilst Boro Saikia et al found evidence of a 14 year solar-like cycle on 61 Cyg A~\citep{Boro2018}.

The magnetic fields of many M-dwarfs appear to be very stable ~\citep{Morin08,Vida16}, although work by~\cite{Lavail2018} suggests that the star AD Leo has long term variability in its magnetic field. A changing in field structure of a star will lead to variations in the latitude of the dipole axis. Stars could move across Figure~\ref{fig:massdistribution} as their dipole tilt varies, showing a compact band of prominence material around the equatorial plane one year but a broader band another year with higher latitude prominences. This is seen by a few stars in this sample for which there are multiple maps. This would allow for a star with a low stellar inclination to present visible prominences in one year whilst this may not have been possible the year before.\\

In Table~\ref{tab:vismassdata} the percentage of prominence mass that would be visible from each map is listed. For many maps there are no visible prominence sites, the exceptions are GJ1156 (2007/2009), EQ Peg A (2006), GJ1111 (2008), GJ9520 (2008), GJ494 (2007) and V374 Peg (2005). In the case of GJ9520, the prominence sites that would be visible are very close to the stellar surface and for this reason not further investigated. These prominences are likely closer to solar prominences than the slingshot prominences investigated here and are unlikely to be observable. For most maps showing visible prominence material, the visible mass is less than 1\% of the total prominence mass supported. V374 Peg is of particular interest, as this is the star on which prominences have also been observed. Here we predict from the 2005 ZDI map that only 13\% of the total prominence mass supported would have been visible to observers. This is an upper limit as this work assumes all prominence sites to be filled at a given time, and thus in reality some of the sites in a visible location may not be filled at the time of observation. These results suggest that observations for such stars could be showing only a very small proportion of the total prominence mass. For stars such as AB Dor or Speedy Mic, where prominences are well studied, the observed masses are likely to be much closer to the total value. This is a reflection of the co-rotation radius of these stars being significantly closer to the stellar surface and making the prominences on this star much more likely to be visible.\\

%\newpage
\section{The importance of prominences}
\subsection{Prominences as mass and angular momentum loss mechanisms}

%--------FIGURE--------
\begin{figure}
    \centering
    \includegraphics[width=1\columnwidth]{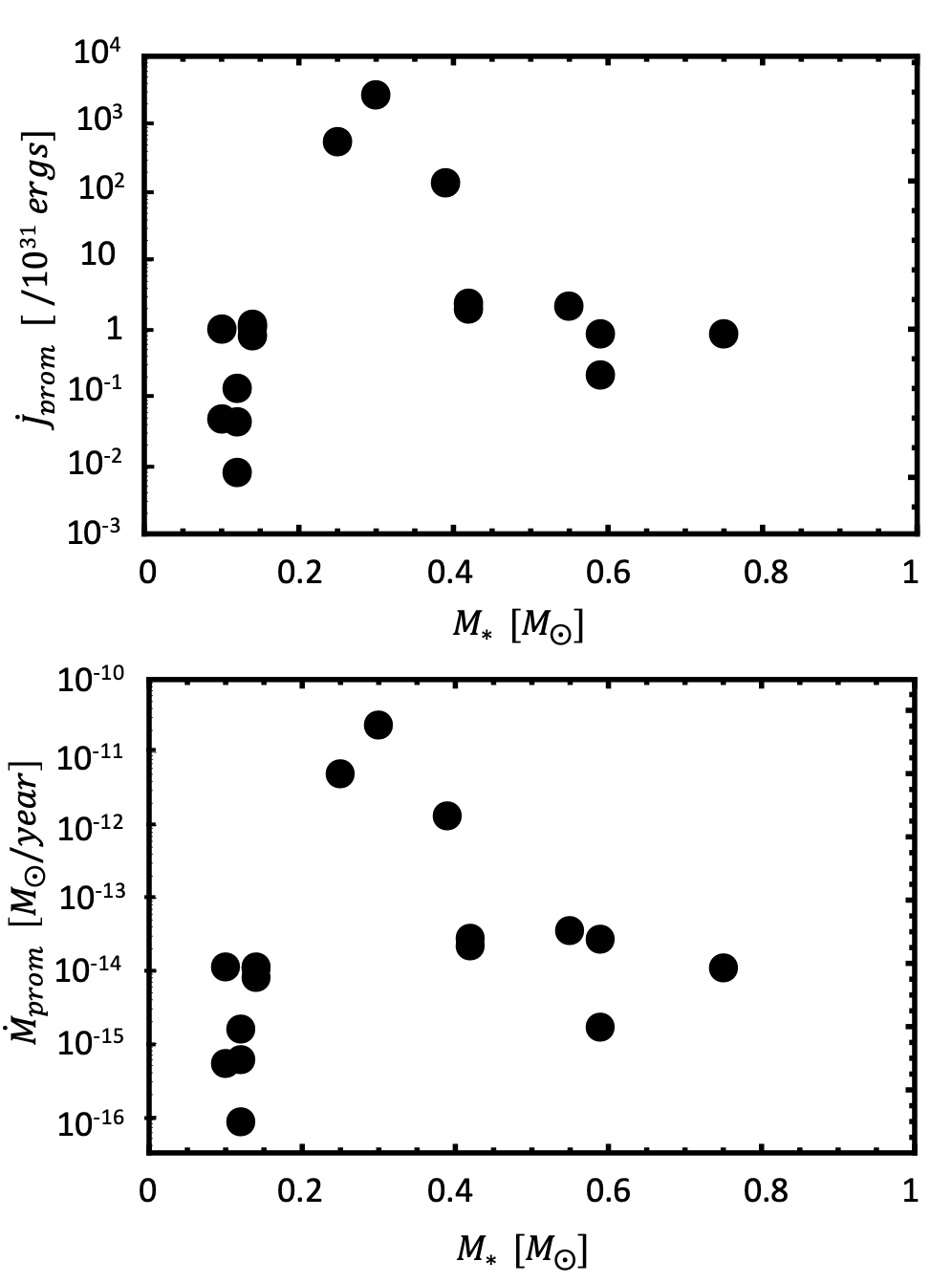}
    \caption{Angular momentum loss rates (top panel) and mass loss rates (bottom panel) for the prominences against stellar mass. The data are given in Table~\ref{tab:spindown}.}
    \label{fig:JdotMdot}
\end{figure}
%---------------------

%------TABLE--------
\begin{table}
\begin{center}
 \begin{tabular}{|c c c c|} 
 \hline
 Map & $\dot{M}_\mathrm{prom}$ & $\dot{J}_\mathrm{prom}$ & Prominence spin-down  \\
 & [M$_{\odot}$/year] & [ergs] & timescale [Gyr] \\
 \hline
 EQ Peg B 2006 & 5.2$\times10^{-12}$ & 5.8$\times10^{33}$& 0.2 \\ 
 GJ1156 2007  & 1.2$\times10^{-14}$ &1.2$\times10^{31}$ & 20.0 \\
 GJ1156 2008 & 8.4$\times10^{-15}$ &8.1$\times10^{30}$ & 29.4 \\
 GJ1156 2009 & 1.1$\times10^{-14}$& 1.1$\times10^{31}$& 21.4 \\
 AD Leo 2007 &2.8$\times10^{-14}$ & 2.4$\times10^{31}$& 24.2 \\
 AD Leo 2008 & 2.2$\times10^{-14}$& 2.0$\times10^{31}$& 29.4 \\
 EQ Peg A 2006 & 1.4$\times10^{-12}$& 1.4$\times10^{33}$& 0.7  \\
 GJ1111 2007  &1.1$\times10^{-14}$& 1.0$\times10^{31}$& 5.5 \\
 GJ1111 2008 & 5.4$\times10^{-16}$& 4.9$\times10^{29}$& 116.7 \\
 GJ1245b 2006  & 6.2$\times10^{-16}$& 4.5$\times10^{29}$& 159.3 \\
 GJ1245b 2007 &1.6$\times10^{-15}$ &1.4$\times10^{30}$ & 52.6 \\
 GJ1245b 2008 & 8.5$\times10^{-17}$ &8.1$\times10^{28}$ & 894.8 \\
 GJ9520 2008 & 3.6$\times10^{-14}$ & 2.2$\times10^{31}$& 38.2 \\
 GJ182 2007  &1.1$\times10^{-14}$& 8.9$\times10^{30}$& 285.6 \\
 GJ494 2007  & 2.8$\times10^{-14}$&8.8$\times10^{30}$& 114.2\\
 GJ494 2008 & 1.7$\times10^{-15}$& 2.2$\times10^{30}$& 588.2 \\
V374 Peg 2006  & 2.4$\times10^{-11}$ &2.8$\times10^{34}$& 0.1 \\
 \hline
\end{tabular}
\caption{Table of mass and angular momentum loss rates. Also shown are the spin down timescales associated with the prominence angular momentum loss rates.}
\label{tab:spindown}
\end{center}
\end{table}
%-----------------

The angular momentum and mass loss rates for the prominences are found and plotted against stellar mass in Figure~\ref{fig:JdotMdot}. There is a trend for mass and angular momentum loss rates to increase as stellar mass decreases. The stars in the bistable regime, which are the very low mass stars in our sample, are the exception here, showing much smaller mass and angular momentum loss rates than would be expected from these trends. The low mass-loss rates from the prominences predicted for these stars are the result of the weak and complex field structures that these stars exhibit. Their weak fields result in low surface mass densities in active regions and the fewer stable points in their coronae lead to a significantly smaller total prominence mass than on other stars. The total prominence mass supported will scale with the field strength squared, which in general increases with decreasing stellar mass.
Meanwhile, the stars showing the largest prominence mass loss rates are V374 Peg, EQ Peg B and EQ Peg A. The reasons for this are two-fold. The first reason is the magnetic field strength and structure, which lead to supporting large prominence masses. Secondly, these stars show some of the smallest co-rotation radii in the stellar sample. Two stars with similar magnetic fields will support different quantities of total prominence mass, depending on their co-rotation radii. Since the magnetic field strength decreases with distance from the star, the star with a smaller co-rotation radius will have a stronger field at this critical radius than the star with the larger co-rotation radius. This larger field strength will enable more prominence material to be supported at the lower co-rotation radius than the larger one, and, as this makes up such a large proportion of the total prominence mass, will mean the star with the low co-rotation radius supports more total prominence material.\\

The angular momentum loss rates reflect a similar pattern to the mass loss, which is unsurprising as these quantities scale linearly. The spin-down timescales for the stars as a consequence of prominences ejection are calculated from the angular momentum loss rates as $t = J_{\star} / \dot{J}_\mathrm{prom}$ and are listed in Table~\ref{tab:spindown}. These spin-down timescales vary across the sample from 0.1Gyrs for V374 Peg to 588.2Gyrs for GJ494 (2008). As the results for $\dot{J}_\mathrm{prom}$ are likely lower limits on the true value, these spin-down timescales will be upper limits. It is interesting that the prominences provide a regulation mechanism, leading to convergence of rotation rates (as the dynamo does in the unsaturated regime) with faster rotators losing more angular momentum.\\

For both the mass and angular momentum loss rates, the range of values found spans around 5 or 6 orders of magnitude whilst the stellar mass only varies by a factor of about 4. There is no single $\dot{M}_\mathrm{prom}$ or $\dot{J}_\mathrm{prom}$ value that could be chosen for this sample, despite the relatively small variation in stellar mass. This relates to the wide range in total prominence mass that can be supported on these stars and is influenced by magnetic field strength, rotation rate and therefore co-rotation radius and the latitude of the dipole axis.

\subsection{Prominences as wind gauges}

\begin{figure}
    \centering
    \includegraphics[width=1\columnwidth]{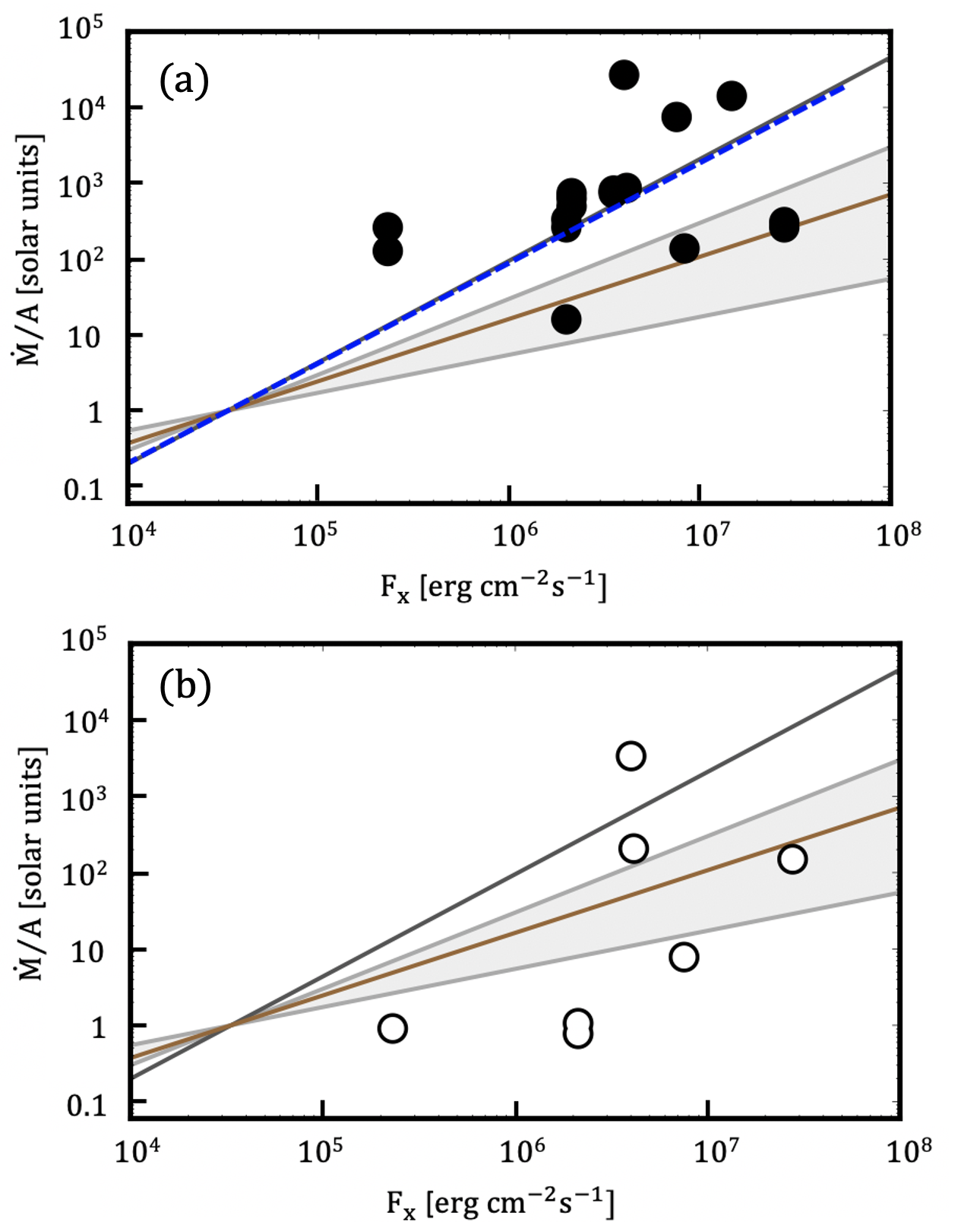}
    \caption{(a) The predicted mass loss rates per unit area for the prominences (black points). The line of best fit for this data is shown by the blue dashed line. The grey shaded region shows the range of wind mass loss rates predicted by Ahuir et al.
    %~\cite{Ahuir2020}
    and dark grey line shows the observationally-based wind mass loss rates from Wood et al.
    %~\cite{Wood2005}.
    The fit from Suzuki et al.
    %~\cite{Suzuki2013}
    for wind mass loss rates is shown in brown. Table~\ref{tab:woodsrelations} shows the corresponding equations for these fits. (b) shows the results using the visible mass. Note that some maps did not show any visible prominences.}
    \label{fig:woodsplot}
\end{figure}

The prominence mass loss rates per unit area can be calculated and compared to published wind models. Figure~\ref{fig:woodsplot} (a) shows this mass loss rate per unit area plotted against the stellar X-ray flux, the values of which were extracted from Vizier~\citep{vizierweb} and Simbad~\citep{simbadweb}, as black points. The X-ray fluxes can be found in the following references;~\cite{Stelzer13,Haakonsen09,Malo13}.
The literature values for wind mass loss rates per unit area, tabulated in Table~\ref{tab:woodsrelations}, are plotted for comparison. The line of best fit is also found for the prominence data ($\dot{M}/A \propto F_X^{1.32}$) and plotted as a blue dashed line. This matches very closely to the wind relation found by~\cite{Wood2005}, providing further evidence that prominences would be good proxies for measuring wind mass loss rates~\citep{WindGauges}. \\

Whilst the line of best fit matches well to the literature, the data show a reasonable amount of scatter. This is also apparent amongst maps taken on consecutive years of the same star, suggesting that the scatter is intrinsic rather than caused by stellar properties such as mass, rotation rate or age. From Figure~\ref{fig:massdistribution} it is clear that the latitude of the dipole axis can vary significantly between years for a given star. This variation determines the magnetic field structure and thus affects the prominence mass that can be supported, translating to the scatter seen in this plot.\\

The prominence mass loss rate for V374 Peg is particularly interesting as it can be compared to the value from~\cite{JardineCC19} based on the observed prominence data. From the observed prominence masses and lifetimes, the authors predicted a mass loss rate per unit area of ~$2\times 10^4$ solar units.
%, by assuming a surface area of 1\% to contribute to the prominence. 
The prominence mass loss rate per unit area found here is 10 times their predicted value, but is consistent with their work as we predict only around 10\% of the total prominence mass to be visible on this map.\\

In Table~\ref{tab:woodsdata}, we list the surface area contribution to the prominences from this work. In the vast majority of the maps used here, we predict the surface area that contributes to the prominence mass loss rates to be very small; only above 1\% for the maps of EQ Peg A, GJ1111 (2007), EQ Peg B and V374 Peg. This suggests that using 1\% in calculations based on the observed data is a reasonable assumption. However, this work has shown that in using prominence mass loss rates to infer the wind mass loss rates results are likely to be biased by the proportion of mass visible to the observer. In underestimating the mass of the observed prominence material, the predicted mass loss rate of the prominence will also be underestimated and thus the wind mass loss rates also. In Table~\ref{tab:vismassdata} we report the percentage of visible prominence mass for each map in this sample which, for those maps that showed any visible mass, is often below 1\%. In (b) of Figure~\ref{fig:woodsplot} we show the prominence mass loss rates per unit area that would be predicted from the visible mass. From this sample of stars it is apparent the extent to which the mass loss rate of stellar winds could be underestimated if the observed prominence mass was used as a measure.

%------TABLE--------
\begin{table}
\begin{center}
 \begin{tabular}{|c c|} 
 \hline
 Relation & Reference\\
 $\dot{M}/A \propto F_X^a$ & \\
 \hline
  $a = 1.34$ &~\cite{Wood2005}\\
  $a = 0.82$ &~\cite{Suzuki2013}\\
  $a = 0.5$ to $1$ &~\cite{Ahuir2020}\\
  $a = 1.32$ & the prominence model in this work\\
 \hline
\end{tabular}
\caption{Table of relations between $\dot{M}/A$ and $F_X$.}
\label{tab:woodsrelations}
\end{center}
\end{table}
%-----------------

%--------FIGURE--------
%---------------------

%------TABLE--------
\begin{table}
\begin{center}
 \begin{tabular}{|c c c c|} 
 \hline
 Map & $F_X$ &  Prominence & Surface area\\
  &  & $\dot{M}/A$ & contributing to\\
  (year) & [$10^6$ ergcm$^{-2}$s$^{-1}$] & [$10^{-14} A_{\odot}$] & prominences [\%]\\
 \hline
 EQ Peg B 2006 & 6.51 & 14103 & 4.67 \\ 
 GJ1156 2007 & 0.46 & 759 & 0.48 \\
 GJ1156 2008 & 0.02 & 624 & 0.42  \\
 GJ1156 2009 & 0.02 & 514 & 0.68 \\
 AD Leo 2007 & 3.52 & 752 & 0.21 \\
 AD Leo 2008 & 3.85 & 816 & 0.15 \\
 EQ Peg A 2006 & 5.09 & 7643 & 1.15 \\
 GJ1111 2007 & 0.06 & 265 & 2.85 \\
 GJ1111 2008 & 0.02 & 128 & 0.28 \\
 GJ1245b 2006 & 0.95 & 327 & 0.07  \\
 GJ1245b 2007 & 0.20 & 267 & 0.25 \\
 GJ1245b 2008 & 4.27 & 17 & 0.21 \\
 GJ9520 2008 & 3.87 & 890 & 0.13 \\
 GJ182 2007 & 1.20 & 135 & 0.10 \\
 GJ494 2007 & 0.34 & 298 & 0.26\\
 GJ494 2008 & 0.01 & 253 & 0.02 \\
 V374 Peg 2006 & 54.93 & 25857 & 6.48\\
 \hline
\end{tabular}
\caption{Table of values for Figure~\ref{fig:woodsplot} (a). X-ray fluxes are calculated from the X-ray fluxes observed at Earth and the distance to each star.}
\label{tab:woodsdata}
\end{center}
\end{table}
%-----------------

\subsection{The mass loss rate as seen by an orbiting planet}

\begin{figure}
    \centering
    \includegraphics[width=1\columnwidth]{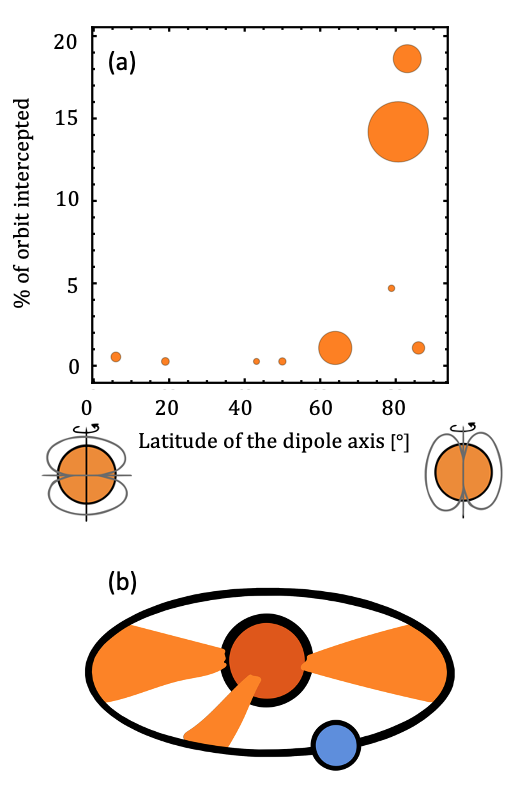}
    \caption{(a) The percentage of a planetary orbit that is intercepted by ejected prominences,  plotted against dipole tilt (right hand side being a dipole axis aligned with the rotation axis and left hand side being perpendicular, as shown by the cartoons). The size of the points represent the maximum $\dot{M}_\mathrm{prom}$ intercepting the planet throughout the orbit. (b) shows a cartoon of the system, where ejected prominences may intercept the path or an orbiting planet.}
    \label{fig:planetmassloss}
\end{figure}

The stellar mass loss has consequences not just for the star, but for any orbiting planets. As a planet orbits the star, it encounters not only the ambient background stellar wind, but also, intermittently,  ejected prominence material. In Figure~\ref{fig:planetmassloss} we plot the percentage of an equatorial planetary orbit for which a planet could intercept ejected prominence material against the latitude of the stellar dipole axis.
%We assume our planet to be within the equatorial plane, and assume that any prominence material that could collide with the orbiting planet if ejected, does. By selecting out the prominence stable points in longitude that are within the latitude band that could collide with the planet, we calculate the angular extent of the orbit that is encompassed by interaction with possible prominence eruptions. In this way, we calculate the maximum percentage of the planetary orbit that could be affected.\\

For almost half of the maps (GJ494 2007, GJ494 2008, GJ1156 2007, GJ1156 2008, GJ9580, GJ1245b 2008, AD Leo 2008) such a planet would not have intercepted any prominence mass from the host star at all, and these are therefore not shown on Figure~\ref{fig:planetmassloss}. For the remaining maps with a nonzero intercepted mass flux, the planet would intercept the prominences for less than a fifth of its orbit, but for many this figure is below 2\%. Planets around stars whose magnetic fields have high dipole latitudes (greater than 60$^\circ$) would be those most likely to have their paths frequently intercepted by large quantities of prominence material. Here the planet was assumed to orbit within the equatorial plane of the star. This is the most common orientation for planetary orbits and from this work it suggests the most likely to be hit by ejected prominences. Prominence material typically gathers about the equatorial plane of the star, as seen in Figures~\ref{fig:massdistribution},~\ref{fig:visibilities1} and~\ref{fig:visibilities2}, and thus planets that orbit here will be the most affected by prominence ejections. Planets that have inclined orbits will experience less ejected prominence material since they will only pass through the equatorial plane in two locations.
%---------------------

%\newpage
\section{Conclusions}
In this paper we have constructed the coronal magnetic field structure for a range of M-dwarfs from their observed ZDI maps and used this to predict prominence formation sites as mechanical stable points within this field geometry. In investigating the locations of these prominence sites, we have found them to be dependent on the alignment of the rotation and magnetic dipole axes. Large misalignments between these axes allow for prominences to form at high latitudes of the star, whilst maps with good alignments of the dipole and rotation axes show prominence formation around the equatorial plane in a narrow band of latitudes. While the inclination of a star's rotation axis is fixed, its dipole axis may change its location  throughout its cycle, and this will affect whether the star hosts any observable prominences.\\

All the magnetic fields investigated here are predicted to support prominences, but many of these are not visible. Several features favour prominence detection:
\begin{itemize}
    \item If a star has a rotation axis with a high inclination to the observer's line of sight, then latitudes close to the equatorial plane are visible. Since the most massive prominences form around the equatorial plane, this would make these prominences easier to see in the H$_{\alpha}$ spectra.\\
    \item Even if the rotation axis is at a low inclination, prominences may still transit if they form at high latitudes. This may occur when the star has a highly tilted dipole axis.\\
    \item Stars with small co-rotation radii will also make the best candidates for hosting observable prominences. The largest prominences gather around the equatorial co-rotation radius and the closer this is to the star the more likely it is that the prominence will cross the stellar disc. For a fixed inclination, the range of latitudes that are visible to an observer drops off steeply with distance from the star. The extent of this can be seen in the visibility plots. Thus, if the prominence forms at low heights then the range of latitudes that would allow this prominence to be observable is greatly increased, and this can compensate for lower stellar inclinations.\\
    \item From this work we predict that V374 Peg, EQ Peg A, EQ Peg B, GJ1156, GJ1111, GJ1245b and GJ494 could host observable prominences.
\end{itemize}

%In practice, for the prominences to be visible in the H$_{\alpha}$ spectra, their lifetimes would also be important. Secondly, in order to be able to pick out prominences in the observed spectra, they must be massive enough to scatter enough light such that the prominence trace is visible over the background noise. They must also cross enough of the stellar disc for it to be clear that the signature is the result of a co-rotating stellar cloud. This paper has not touched on the lifetimes of these prominences, nor if they would be stable for long enough to be visible observationally.\\

We calculate the mass and angular momentum loss rates for the stars in our sample and plotting against stellar mass show two distinct categories; the very low mass stars that fall into the ``bistable'' regime and the higher mass M-dwarfs. The lowest mass stars (around $0.1M_{\odot}$) with weak and complex magnetic fields support significantly less prominence mass and therefore show much lower prominence mass loss rates than other stars.  For the higher mass stars in the sample, those with lower stellar masses are those that support the highest prominence masses and therefore show largest prominence mass loss rate. The angular momentum loss rates follow the same trend as the mass loss rates. The spin-down timescales due to the prominences are estimated from the angular momentum loss rates ($J_{\star}/\dot{J}_\mathrm{prom}$) with values ranging between 0.1-588 Gigayears. We note two things: firstly, that the prominence mass loss rate for a star could vary as the star progresses through its cycle, as maps with more inclined fields typically support less mass than aligned ones.  Secondly, the mass loss rates here are upper limits, as they assume that all of the predicted prominence support sites are filled.\\
%when in reality it is likely that some will be empty - since the sites repeatedly fill up and are ejected in a limit cycle.\\

It is also worth noting here that for stars that are viewed almost pole on, prominences may be viewed in emission for example LQ Lup and V830Per \citep{2000MNRAS.316..699D,2015MNRAS.453.3706D}. Here we have investigated only the visibility of prominences in terms of the absorption features and not in emission. The advantage of these stars is that they provide an ideal opportunity to estimate the entire prominence mass observationally, and therefore mass loss and angular momentum loss rates, as prominence material is not blocked from view by the stellar disc.\\

We plot $\dot{M}/A$ from our synthetic prominence data against X-ray flux, produce a line of best fit for our data and compare to literature values for wind models. We note that there is quite a large scatter in the data and that this scatter is also present for stars for which we have multiple maps. This suggests the scatter to be intrinsic. Despite the scatter, our line of best fit yields $\dot{M}/A \propto$ $F_X^{1.32}$ which is consistent with the literature and particularly close to the value calculated by~\cite{Wood2005}. This result agrees with the work by~\cite{WindGauges} that suggested that prominences could be a possible way of estimating the wind mass loss rates, since the winds for these stars are so difficult to measure. The surface area contributing to the prominences for the maps used here is generally < 1\%, suggesting the area estimated by~\cite{WindGauges} in their work (1\%) to be reasonable. Within our sample however, V374 Peg is the only star for which prominences have been observed. For the magnetic map we have of this star we predict the visible mass to be a factor of 10 less than the total mass it supports. The $\dot{M}/A$ predicted by~\cite{WindGauges} from the observations is a factor of 10 less than we find here and we suggest this factor of 10 to be due to the underestimation of the observed masses. In using the prominences to predict the wind mass loss rates, we must be careful that the prominence mass we are observing around stars such as these is likely to be an underestimation. This will depend however on the stellar inclination, latitude of the dipole axis and the location of the co-rotation radius.\\

For a planet in an equatorial orbit around the stars in our sample, the prominence mass loss would be intercepting the planet for typically <2\% of the orbit, if at all. EQ Peg B and V374 Peg show much greater fractions of the planetary orbit intercepted by prominence material (18\% and 14\% respectively). Increasing the latitude of the dipole axis increases this fraction. This is because these maps have their dipole axis most closely aligned with the rotation axis and thus can form more, and larger, prominences around their equators than maps with lower dipole latitudes.\\

In conclusion, we have shown that the strength and geometry of stars' magnetic fields have a significant impact on the mass and visibility of prominences that may be supported. Of the M-dwarfs in our sample, the highest mass stars, which tend to have strong and simple fields are the most promising hosts for prominences, while those with the weakest and most complex fields host much lower masses of prominences. Despite the higher-mass M-dwarfs supporting a large quantity of prominence mass, which corresponds to mass loss rates as high as $10^{-11}$M$_{\odot}$/year, much of this mass may not be geometrically visible to observers, particularly if the stars' co-rotation radii are large. Rapidly-rotating stars with small co-rotation radii, rotation axes at high inclinations and aligned magnetic fields make the most promising candidates for prominence detection.  We show that prominences could be used as wind gauges as suggested by~\cite{WindGauges} but that values calculated from observed data could greatly underestimate the result.

\section*{Data Availability}

%The data used to make Figures \ref{fig:slope}, \ref{fig:excess} and \ref{fig:confusagram} are in Table \ref{table:StellarSample}  and can be accessed at  \url{https://doi.org/10.17630/4dea9f0c-9626-4655-ba61-8af5c05fbe38}. 
The research data listed in the tables here can be accessed at \url{https://doi.org/10.17630/c9a3905e-c66e-42fc-aede-202c2fa4306b}~\citep{data}.
Archival data underpinning the magnetograms used in this paper is available at polarbase (\url{http://polarbase.irap.omp.eu}). 

\section*{Acknowledgements}
The authors acknowledge support from STFC consolidated grant number ST/R000824/1. The authors would like to thank the anonymous referee for their helpful suggestions that improved the paper.

%%%%%%%%%%%%%%%%%%%%%%%%%%%%%%%%%%%%%%%%%%%%%%%%%%



\bibliographystyle{mnras}
\bibliography{MAIN}

\begin{thebibliography}{}
\makeatletter
\relax
\def\mn@urlcharsother{\let\do\@makeother \do\$\do\&\do\#\do\^\do\_\do\%\do\~}
\def\mn@doi{\begingroup\mn@urlcharsother \@ifnextchar [ {\mn@doi@}
  {\mn@doi@[]}}
\def\mn@doi@[#1]#2{\def\@tempa{#1}\ifx\@tempa\@empty \href
  {http://dx.doi.org/#2} {doi:#2}\else \href {http://dx.doi.org/#2} {#1}\fi
  \endgroup}
\def\mn@eprint#1#2{\mn@eprint@#1:#2::\@nil}
\def\mn@eprint@arXiv#1{\href {http://arxiv.org/abs/#1} {{\tt arXiv:#1}}}
\def\mn@eprint@dblp#1{\href {http://dblp.uni-trier.de/rec/bibtex/#1.xml}
  {dblp:#1}}
\def\mn@eprint@#1:#2:#3:#4\@nil{\def\@tempa {#1}\def\@tempb {#2}\def\@tempc
  {#3}\ifx \@tempc \@empty \let \@tempc \@tempb \let \@tempb \@tempa \fi \ifx
  \@tempb \@empty \def\@tempb {arXiv}\fi \@ifundefined
  {mn@eprint@\@tempb}{\@tempb:\@tempc}{\expandafter \expandafter \csname
  mn@eprint@\@tempb\endcsname \expandafter{\@tempc}}}

\bibitem[\protect\citeauthoryear{Ahuir, Brun  \& Strugarek}{Ahuir
  et~al.}{2020}]{Ahuir2020}
Ahuir J.,  Brun A.,   Strugarek A.,  2020, \mn@doi [\aap]
  {10.1051/0004-6361/201936974}, 635, A170

\bibitem[\protect\citeauthoryear{Barnes, Cameron, Unruh, Donati  \&
  Hussain}{Barnes et~al.}{1998}]{Barnes98}
Barnes J.~R.,  Cameron A.~C.,  Unruh Y.~C.,  Donati J.-F.,   Hussain G. A.~J.,
  1998, \mn@doi [\mnras] {10.1046/j.1365-8711.1998.01805.x}, 299, 904

\bibitem[\protect\citeauthoryear{Barnes, Cameron, James  \& Donati}{Barnes
  et~al.}{2000}]{Barnes2000}
Barnes J.~R.,  Cameron A.~C.,  James D.~J.,   Donati J.-F.,  2000, \mn@doi
  [\mnras] {10.1046/j.1365-8711.2000.03237.x}, 314, 162

\bibitem[\protect\citeauthoryear{Bochanski, Hawley, K.~R.~Covey, Reid,
  Golimowski  \& Ivezic}{Bochanski et~al.}{2010}]{Bochanski2010}
Bochanski J.~J.,  Hawley S.~L.,  K.~R.~Covey A. A.~W.,  Reid I.~N.,  Golimowski
  D.~A.,   Ivezic Z.,  2010, \mn@doi [\aj] {10.1088/0004-6256/143/6/152}, 139,
  2679

\bibitem[\protect\citeauthoryear{Byrne, Eibe  \& Rolleston}{Byrne
  et~al.}{1996}]{ByrneEibe1996}
Byrne P.~B.,  Eibe M.~T.,   Rolleston W. R.~J.,  1996, \aap, 311, 651

\bibitem[\protect\citeauthoryear{CDS-Strasbourg}{CDS-Strasbourg}{2020b}]{simbadweb}
CDS-Strasbourg 2020b, SIMBAD Astronomical Database, \url
  {http://simbad.u-strasbg.fr/simbad/}

\bibitem[\protect\citeauthoryear{CDS-Strasbourg}{CDS-Strasbourg}{2020a}]{vizierweb}
CDS-Strasbourg 2020a, VizieR, \url {https://vizier.u-strasbg.fr/viz-bin/VizieR}

\bibitem[\protect\citeauthoryear{Cameron}{Cameron}{1999}]{ACStellProm}
Cameron A.~C.,  1999, \aspc, 158, 146

\bibitem[\protect\citeauthoryear{Cameron \& Robinson}{Cameron \&
  Robinson}{1989}]{CC1989}
Cameron A.~C.,  Robinson R.,  1989, \mn@doi [\mnras] {10.1093/mnras/236.1.57},
  236, 57

\bibitem[\protect\citeauthoryear{Cameron \& Woods}{Cameron \&
  Woods}{1992}]{CCWoods92}
Cameron A.~C.,  Woods J.~A.,  1992, \mn@doi [\mnras] {10.1093/mnras/258.2.360},
  258 (2), 360

\bibitem[\protect\citeauthoryear{Cameron, D.~K.~Duncan, Foing, Kuntz, Pension,
  Robinson  \& Soderblom}{Cameron et~al.}{1990}]{CC1990}
Cameron A.~C.,  D.~K.~Duncan P.~E.,  Foing B.~H.,  Kuntz K.~D.,  Pension M.~V.,
   Robinson R.~D.,   Soderblom D.~R.,  1990, \mn@doi [\mnras]
  {10.1093/mnrasl/slx206}, 247, 415

\bibitem[\protect\citeauthoryear{Cameron et~al.,}{Cameron
  et~al.}{1999}]{Cameron1999}
Cameron A.~C.,  et~al., 1999, \mn@doi [\mnras]
  {10.1046/j.1365-8711.1999.02771.x}, 308, 493

\bibitem[\protect\citeauthoryear{{Donati}, {Mengel}, {Carter}, {Marsden},
  {Collier Cameron}  \& {Wichmann}}{{Donati}
  et~al.}{2000}]{2000MNRAS.316..699D}
{Donati} J.-F.,  {Mengel} M.,  {Carter} B.~D.,  {Marsden} S.,  {Collier
  Cameron} A.,   {Wichmann} R.,  2000, \mn@doi [\mnras]
  {10.1046/j.1365-8711.2000.03570.x}, \href
  {http://adsabs.harvard.edu/abs/2000MNRAS.316..699D} {316, 699}

\bibitem[\protect\citeauthoryear{Donati et~al.,}{Donati
  et~al.}{2008}]{Donati08}
Donati J.-F.,  et~al., 2008, \mn@doi [\mnras]
  {10.1111/j.1365-2966.2008.13799.x}, 390(2), 545

\bibitem[\protect\citeauthoryear{{Donati} et~al.,}{{Donati}
  et~al.}{2015}]{2015MNRAS.453.3706D}
{Donati} J.~F.,  et~al., 2015, \mn@doi [\mnras] {10.1093/mnras/stv1837}, \href
  {https://ui.adsabs.harvard.edu/abs/2015MNRAS.453.3706D} {453, 3706}

\bibitem[\protect\citeauthoryear{Dunstone, Barnes, Cameron  \&
  Jardine}{Dunstone et~al.}{2006a}]{Dunstone2005}
Dunstone N.~J.,  Barnes J.~R.,  Cameron A.~C.,   Jardine M.,  2006a, \mn@doi
  [\mnras] {10.1111/j.1365-2966.2005.09729.x}, 365, 530

\bibitem[\protect\citeauthoryear{Dunstone, Cameron, Barnes  \&
  Jardine}{Dunstone et~al.}{2006b}]{DunstoneII}
Dunstone N.~J.,  Cameron A.~C.,  Barnes J.~R.,   Jardine M.,  2006b, \mn@doi
  [\mnras] {10.1111/j.1365-2966.2006.11128.x}, 373, 1308

\bibitem[\protect\citeauthoryear{Eibe}{Eibe}{1998}]{Eibe98}
Eibe M.,  1998, \aap, 337, 757

\bibitem[\protect\citeauthoryear{Ferreira}{Ferreira}{2000}]{Ferreira2000}
Ferreira M.,  2000, \mn@doi [\mnras] {10.1046/j.1365-8711.2000.03540.x}, 316,
  647

\bibitem[\protect\citeauthoryear{Guirado et~al.,}{Guirado
  et~al.}{2010}]{Guirado2010}
Guirado J.~C.,  et~al., 2010, ASSSP, Highlights of Spanish Astrophysics V, 139

\bibitem[\protect\citeauthoryear{Haakonsen \& Rutledge}{Haakonsen \&
  Rutledge}{2009}]{Haakonsen09}
Haakonsen C.,  Rutledge R.,  2009, \mn@doi [The Astrophysical Journal
  Supplement Series] {10.1088/0067-0049/184/1/138}, (), 1

\bibitem[\protect\citeauthoryear{Haisch, Linsky, Bornmann, Stencel, Antiochos,
  Golub  \& Vaiana}{Haisch et~al.}{1983}]{Haisch83}
Haisch B.,  Linsky J.,  Bornmann P.,  Stencel R.,  Antiochos S.,  Golub L.,
  Vaiana G.,  1983, \mn@doi [\apj] {10.1086/160866}, 267, 280

\bibitem[\protect\citeauthoryear{Hussain}{Hussain}{2013}]{Hussain2013}
Hussain G. A.~J.,  2013, Proceedings IAU Symposium, No. 300, 309

\bibitem[\protect\citeauthoryear{Innis, Thompson, Coates  \& {Lloyd
  Evans}}{Innis et~al.}{1988}]{Innis1988}
Innis J.~L.,  Thompson K.,  Coates D.~W.,   {Lloyd Evans} T.,  1988, \mn@doi
  [\mnras] {10.1093/mnras/235.4.1411}, 235, 1411

\bibitem[\protect\citeauthoryear{Jakosky et~al.,}{Jakosky
  et~al.}{2015}]{Mars2015}
Jakosky B.~M.,  et~al., 2015, \mn@doi [Science] {10.1126/science.aad0210}, 350
  (6261), aad0210

\bibitem[\protect\citeauthoryear{Jardine \& Cameron}{Jardine \&
  Cameron}{1991}]{JardineCC91}
Jardine M.,  Cameron A.~C.,  1991, \solphys, 131, 269

\bibitem[\protect\citeauthoryear{Jardine \& Cameron}{Jardine \&
  Cameron}{2019}]{WindGauges}
Jardine M.,  Cameron A.~C.,  2019, \mn@doi [MNRAS] {10.1093/mnras/sty2872}, 482
  (3), 2853

\bibitem[\protect\citeauthoryear{Jardine \& van Ballegooijen}{Jardine \& van
  Ballegooijen}{2005}]{JardineB05}
Jardine M.,  van Ballegooijen A.,  2005, \mnras, 361, 1173

\bibitem[\protect\citeauthoryear{Jardine, Cameron, Donati  \& Hussain}{Jardine
  et~al.}{2020a}]{Jardine19}
Jardine M.,  Cameron A.~C.,  Donati J.-F.,   Hussain G.,  2020a, \mn@doi
  [\mnras] {10.1093/mnras/stz3173}, 491 (3), 4076

\bibitem[\protect\citeauthoryear{Jardine, Cameron, Donati  \& Hussain}{Jardine
  et~al.}{2020b}]{JardineCC19}
Jardine M.,  Cameron A.,  Donati J.-.,   Hussain G.,  2020b, \mn@doi [\mnras]
  {10.1093/mnras/stz3173}, 491(3), 4076

\bibitem[\protect\citeauthoryear{Jeffers et~al.,}{Jeffers
  et~al.}{2018}]{Jeffers18}
Jeffers S.~V.,  et~al., 2018, \mn@doi [\mnras] {10.1093/mnras/sty1717}, 479
  (4), 5266

\bibitem[\protect\citeauthoryear{Jeffries}{Jeffries}{1993}]{Jeffries93}
Jeffries R.~D.,  1993, \mn@doi [\mnras] {10.1093/mnras/262.2.369}, 262, 369

\bibitem[\protect\citeauthoryear{{Khodachenko} et~al.,}{{Khodachenko}
  et~al.}{2007}]{Khodachenko}
{Khodachenko} M.~L.,  et~al., 2007, \mn@doi [Astrobiology]
  {10.1089/ast.2006.0127}, 7, 167

\bibitem[\protect\citeauthoryear{{Korhonen}, {Vida}, {Leitzinger}, {Odert}  \&
  {Kov{\'a}cs}}{{Korhonen} et~al.}{2017}]{2017IAUS..328..198K}
{Korhonen} H.,  {Vida} K.,  {Leitzinger} M.,  {Odert} P.,   {Kov{\'a}cs} O.~E.,
   2017, in {Nandy} D.,  {Valio} A.,   {Petit} P.,  eds,  IAU Symposium Vol.
  328, Living Around Active Stars. pp 198--203 (\mn@eprint {arXiv}
  {1612.06643}), \mn@doi{10.1017/S1743921317003969}

\bibitem[\protect\citeauthoryear{Lavail, Kochukhov  \& Wade}{Lavail
  et~al.}{2018}]{Lavail2018}
Lavail A.,  Kochukhov O.,   Wade G.,  2018, \mn@doi [\mnras]
  {10.1093/mnras/sty1825}, 479 (4), 4836

\bibitem[\protect\citeauthoryear{{Leitzinger}, {Odert}, {Greimel}, {Korhonen},
  {Guenther}, {Hanslmeier}, {Lammer}  \& {Khodachenko}}{{Leitzinger}
  et~al.}{2014}]{2014MNRAS.443..898L}
{Leitzinger} M.,  {Odert} P.,  {Greimel} R.,  {Korhonen} H.,  {Guenther} E.~W.,
   {Hanslmeier} A.,  {Lammer} H.,   {Khodachenko} M.~L.,  2014, \mn@doi
  [\mnras] {10.1093/mnras/stu1161}, \href
  {https://ui.adsabs.harvard.edu/abs/2014MNRAS.443..898L} {443, 898}

\bibitem[\protect\citeauthoryear{Leitzinger, Odert, Zaqarashvili, Hanslmeier
  \& Lammer}{Leitzinger et~al.}{2016}]{Leitzinger2016}
Leitzinger M.,  Odert P.,  Zaqarashvili T.~V.,  Hanslmeier R. G.~A.,   Lammer
  H.,  2016, \mn@doi [\mnras] {10.1093/mnras/stw1922}, 463 (1), 965

\bibitem[\protect\citeauthoryear{Malo, Lafrenière, Artigau, Gagné, F.~Baron
  \& Riedel}{Malo et~al.}{2013}]{Malo13}
Malo L.,  Lafrenière R. D.~D.,  Artigau E.,  Gagné J.,  F.~Baron F.,   Riedel
  A.,  2013, \mn@doi [ApJ] {10.1088/0004-637X/762/2/88}, (), 1

\bibitem[\protect\citeauthoryear{Morin et~al.,}{Morin et~al.}{2008}]{Morin08}
Morin J.,  et~al., 2008, \mn@doi [\mnras] {10.1111/j.1365-2966.2008.13809.x},
  390(2), 567

\bibitem[\protect\citeauthoryear{Morin, Donati, Delfosse, Forveille  \&
  Jardine}{Morin et~al.}{2010}]{Morin10}
Morin J.,  Donati J.~F.,  Delfosse P. P.~X.,  Forveille T.,   Jardine M.~M.,
  2010, \mn@doi [\mnras] {10.1111/j.1365-2966.2010.17101.x}, 407(4), 2269

\bibitem[\protect\citeauthoryear{Morin, Dormy, Schrinner  \& Donati}{Morin
  et~al.}{2011}]{Morin11}
Morin J.,  Dormy E.,  Schrinner M.,   Donati J.,  2011, \mn@doi [\mnras]
  {10.1111/j.1745-3933.2011.01159.x}, 418, 133

\bibitem[\protect\citeauthoryear{Odert, Leitzinger, Guenther  \& Heinzel}{Odert
  et~al.}{2020}]{Odert2020}
Odert P.,  Leitzinger M.,  Guenther E.,   Heinzel P.,  2020, \mn@doi [\mnras]
  {10.1093/mnras/staa1021}, 494 (3), 3766

\bibitem[\protect\citeauthoryear{Reville, Brun, Matt, Bouvier, Folsom  \&
  P.Petit}{Reville et~al.}{2015}]{Reville15}
Reville V.,  Brun A.~S.,  Matt A. S.~S.,  Bouvier J.,  Folsom C.,   P.Petit
  2015, \mn@doi [ApJ] {10.1088/0004-637X/814/2/99}, 814 (2), 99

\bibitem[\protect\citeauthoryear{Saikia et~al.,}{Saikia
  et~al.}{2018}]{Boro2018}
Saikia S.~B.,  et~al., 2018, \mn@doi [\aap] {10.1051/0004-6361/201834347}, 620,
  L11

\bibitem[\protect\citeauthoryear{See et~al.,}{See et~al.}{2018}]{See18}
See V.,  et~al., 2018, \mn@doi [MNRAS] {10.1093/mnras/stx2599}, 474, 536

\bibitem[\protect\citeauthoryear{Shields, Ballard  \& Johnson}{Shields
  et~al.}{2016}]{shields2017}
Shields A.~L.,  Ballard S.,   Johnson J.~A.,  2016, \mn@doi [Phys. Rep.]
  {10.1016/j.physrep.2016.10.003}, 663, 1

\bibitem[\protect\citeauthoryear{Skelly, Unruh, Cameron, Barnes, Donati, Lawson
   \& Carter}{Skelly et~al.}{2008}]{Skelly2008}
Skelly M.,  Unruh Y.,  Cameron A.~C.,  Barnes J.,  Donati J.-F.,  Lawson W.,
  Carter B.,  2008, \mn@doi [\mnras] {10.1111/j.1365-2966.2008.12917.x}, 385,
  708–

\bibitem[\protect\citeauthoryear{Skelly, Unruh, Barnes, Lawson, Donati  \&
  Cameron}{Skelly et~al.}{2009}]{Skelly2009}
Skelly M.~B.,  Unruh Y.~C.,  Barnes J.~R.,  Lawson W.~A.,  Donati J.-F.,
  Cameron A.~C.,  2009, \mn@doi [\mnras] {10.1111/j.1365-2966.2009.15411.x},
  399, 1829

\bibitem[\protect\citeauthoryear{Skelly, Donati, Bouvier, Grankin, Unruh,
  Artemenko  \& Petrov}{Skelly et~al.}{2010}]{Skelly2010}
Skelly M.~B.,  Donati J.-F.,  Bouvier J.,  Grankin K.~N.,  Unruh Y.~C.,
  Artemenko S.~A.,   Petrov P.,  2010, \mn@doi [\mnras]
  {10.1111/j.1365-2966.2009.16132.x}, 403, 159

\bibitem[\protect\citeauthoryear{Stauffer et~al.,}{Stauffer
  et~al.}{2017}]{Stauffer17}
Stauffer J.,  et~al., 2017, \mn@doi [\aj] {10.3847/1538-3881/aa5eb9}, 153, 152

\bibitem[\protect\citeauthoryear{Stelzer, Marino, Micela, Lopez-Santiago  \&
  Liefke}{Stelzer et~al.}{2013}]{Stelzer13}
Stelzer B.,  Marino A.,  Micela G.,  Lopez-Santiago J.,   Liefke C.,  2013,
  \mn@doi [MNRAS] {10.1093/mnras/stt225}, 431 (3), 2063–2079

\bibitem[\protect\citeauthoryear{Suzuki, Imada, Kataoka, Kato, Matsumoto,
  Miyahara  \& Tsuneta}{Suzuki et~al.}{2013}]{Suzuki2013}
Suzuki T.,  Imada S.,  Kataoka R.,  Kato Y.,  Matsumoto T.,  Miyahara H.,
  Tsuneta S.,  2013, \mn@doi [Publications of the Astronomical Society of
  Japan] {10.1093/pasj/65.5.98}, 65(5), id98

\bibitem[\protect\citeauthoryear{Vida et~al.,}{Vida et~al.}{2016}]{Vida16}
Vida K.,  et~al., 2016, \mn@doi [\aap] {10.1051/0004-6361/201527925}, 590, 1

\bibitem[\protect\citeauthoryear{{Vida}, {Leitzinger}, {Kriskovics}, {Seli},
  {Odert}, {Kov{\'a}cs}, {Korhonen}  \& {van Driel-Gesztelyi}}{{Vida}
  et~al.}{2019}]{2019A&A...623A..49V}
{Vida} K.,  {Leitzinger} M.,  {Kriskovics} L.,  {Seli} B.,  {Odert} P.,
  {Kov{\'a}cs} O.~E.,  {Korhonen} H.,   {van Driel-Gesztelyi} L.,  2019,
  \mn@doi [\aap] {10.1051/0004-6361/201834264}, \href
  {https://ui.adsabs.harvard.edu/abs/2019A&A...623A..49V} {623, A49}

\bibitem[\protect\citeauthoryear{Vidotto, Jardine, Opher, Donati  \&
  Gombosi}{Vidotto et~al.}{2011}]{Vidotto11}
Vidotto A.~A.,  Jardine M.,  Opher M.,  Donati J.~F.,   Gombosi T.~I.,  2011,
  Astronomical Society of the Pacific Conference Series, 16th Cambridge
  Workshop on Cool Stars, Stellar Systems, and the Sun., 448, 1293

\bibitem[\protect\citeauthoryear{{Villarreal D'Angelo}, Jardine  \&
  See}{{Villarreal D'Angelo} et~al.}{2018}]{VdA18}
{Villarreal D'Angelo} C.,  Jardine M.,   See V.,  2018, \mnras, 475, 25

\bibitem[\protect\citeauthoryear{{Villarreal D'Angelo}, Jardine, Johnstone  \&
  See}{{Villarreal D'Angelo} et~al.}{2019}]{VdA19}
{Villarreal D'Angelo} C.,  Jardine M.,  Johnstone C.~P.,   See V.,  2019,
  \mn@doi [\mnras] {10.1093/mnras/stz477}, 485 (1), 1448

\bibitem[\protect\citeauthoryear{Waugh \& Jardine}{Waugh \&
  Jardine}{2018}]{Waugh18}
Waugh R.,  Jardine M.,  2018, \mn@doi [\mnras] {10.1093/mnras/sty3225}, 483(2),
  1513

\bibitem[\protect\citeauthoryear{Waugh \& Jardine}{Waugh \&
  Jardine}{2021}]{data}
Waugh R.,  Jardine M.,  2021, \mn@doi []
  {10.17630/c9a3905e-c66e-42fc-aede-202c2fa4306b}

\bibitem[\protect\citeauthoryear{Weber \& Davis}{Weber \&
  Davis}{1967}]{WeberDavis67}
Weber E.,  Davis L.,  1967, \apj, 148, 217

\bibitem[\protect\citeauthoryear{Wood, Muller, Zank, Linsky  \& Redfield}{Wood
  et~al.}{2005}]{Wood2005}
Wood B.~E.,  Muller H.~R.,  Zank G.~P.,  Linsky J.~L.,   Redfield S.,  2005,
  \mn@doi [\apj] {10.1086/432716}, 628(2), L143

\bibitem[\protect\citeauthoryear{van Ballegooijen, Cartledge  \& Priest}{van
  Ballegooijen et~al.}{1998}]{Ballegooijen98}
van Ballegooijen A.,  Cartledge N.~P.,   Priest E.~R.,  1998, \mn@doi [ApJ]
  {10.1086/305823}, 501, 866

\makeatother
\end{thebibliography}
% if your bibtex file is called example.bib

%\begin{thebibliography}{99}
%\end{thebibliography}

%%%%%%%%%%%%%%%%%%%%%%%%%%%%%%%%%%%%%%%%%%%%%%%%%%

%%%%%%%%%%%%%%%%% APPENDICES %%%%%%%%%%%%%%%%%%%%%

%\newpage
\appendix

\section{Coordinate Transform}
\label{app:coordtrans}

\begin{figure}
\centering
		\includegraphics[width=1\columnwidth]{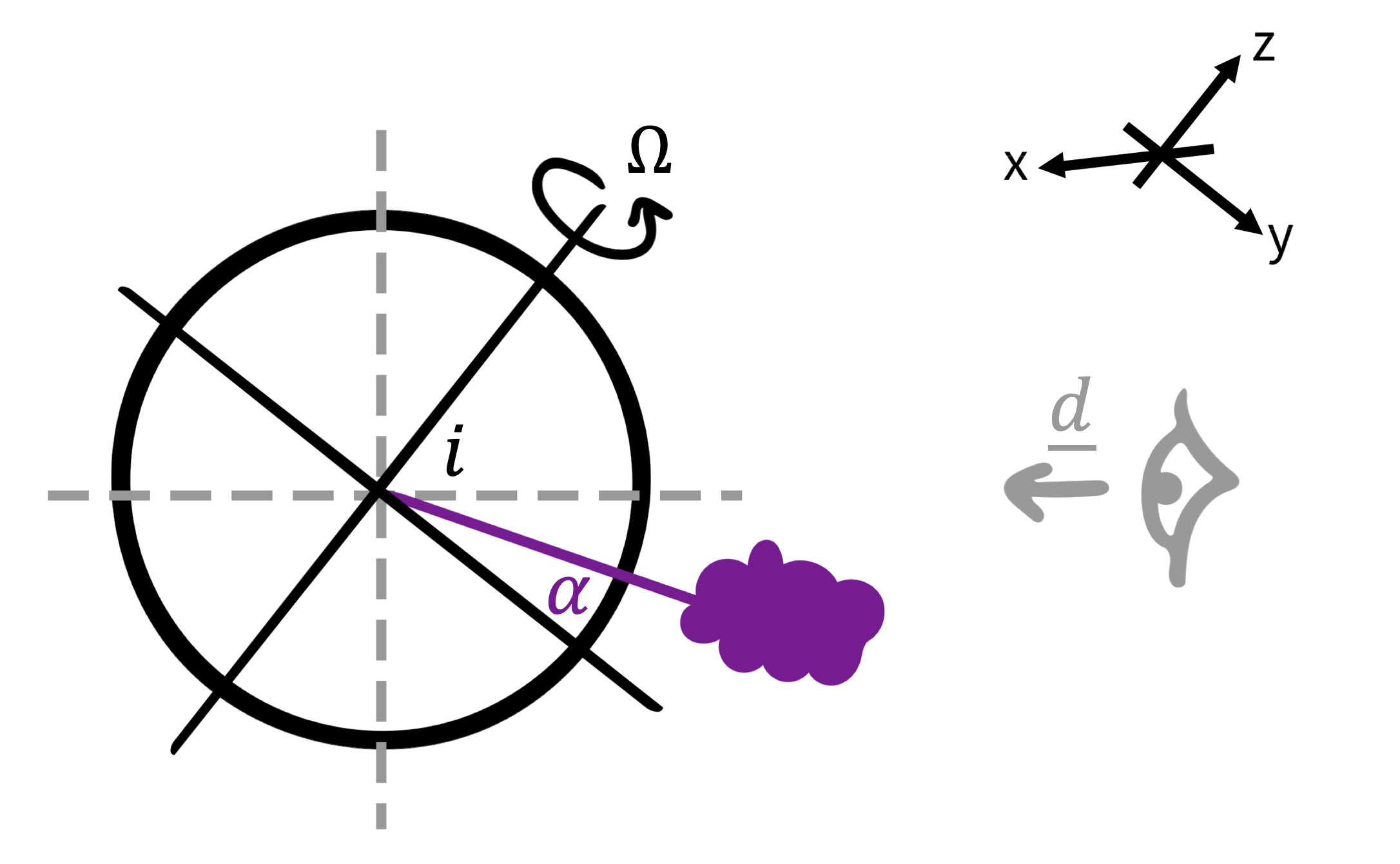}
	\caption{Cartoon showing the scenario of an inclined star with a co-rotating prominence. The important angles of inclination ($i$) and prominence latitude ($\alpha$) and shown.}
    \label{fig:coord1}
\end{figure}

\begin{figure}
\centering
		\includegraphics[width=1\columnwidth]{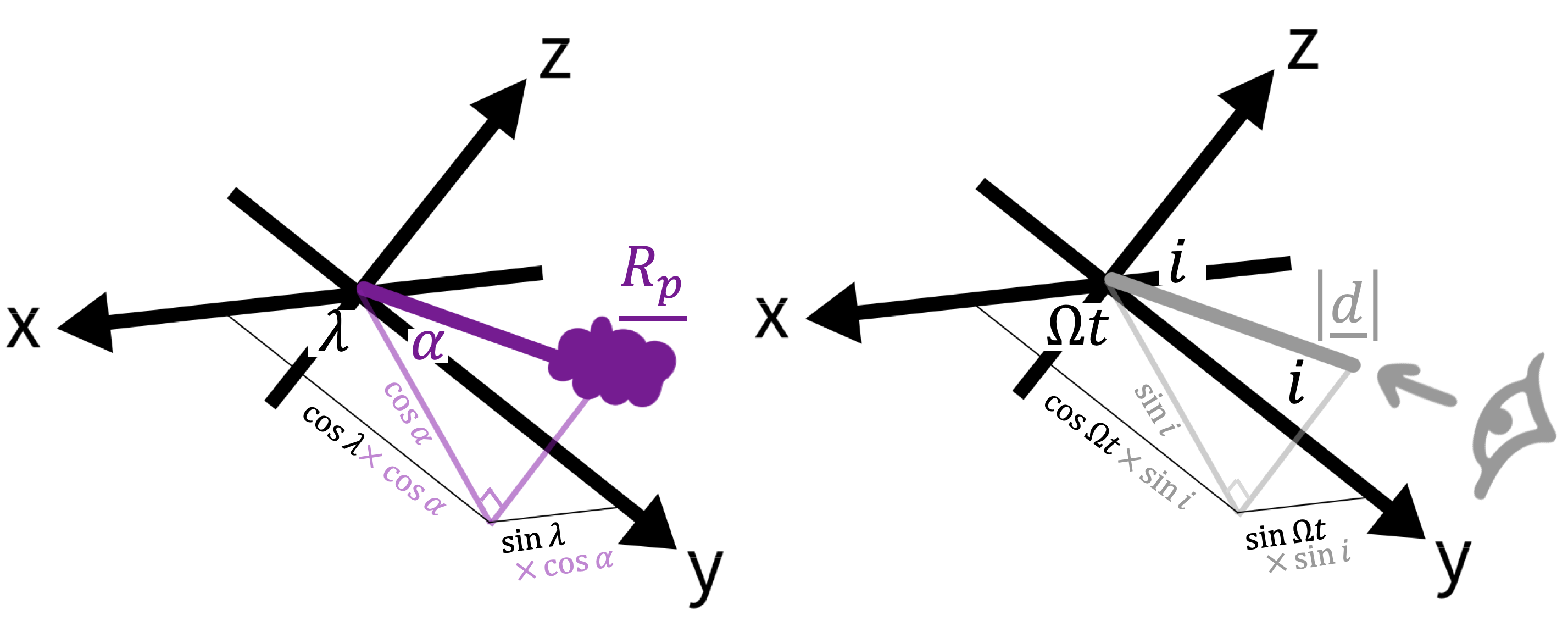}
	\caption{Cartoon for the coordinate transforms for the prominence position ($\underline{R_p}$) on the left and the line of sight vector ($\underline{d}$) on the right.}
    \label{fig:coord2}
\end{figure}

Here we show the coordinate transform used in Section~\ref{sec:vis}.\\ Figure~\ref{fig:coord1} depicts the scenario of a co-rotating prominence and an inclined star.\\
In Figure~\ref{fig:coord2}, we look first at the prominence position, $\underline{R_p}$. The $z$ component is the most simple, and through trigonometry can be seen from the diagram to be $z_p = |\underline{R_p}|\sin{\alpha}$. For the $x$ and $y$ components we use similar trigonometric arguments, though this time the angle in question is the prominence phase $\lambda$ and the hypotenuse of this triangle in the $x-y$ plane is $\cos{\alpha}$. Thus, these components are $x_p = |\underline{R_p}|\cos{\lambda} \times \cos{\alpha}$ and $y_p = |\underline{R_p}|\sin{\lambda} \times \cos{\alpha}$. Combining this all together yields
\begin{equation}
    \underline{R_p} = |\underline{R_p}|(\cos{\lambda}\cos{\alpha}, \sin{\lambda} \cos{\alpha}, \sin{\alpha}).
\end{equation}\\

The same arguments can be made for the line of sight vector, $|\underline{d}|$, though this time the known angle ($i$) is the other angle in the triangle. Thus,  $d_z = |\underline{d}|\cos{i}$, $d_x = |\underline{d}|\cos{\Omega t} \times \sin{i}$ and $y_p = |\underline{d}|\cos{\Omega t} \times \sin{i}$. Since $\underline{d}$ is a unit vector, it's magnitude is 1. We also choose to use the convention of observers in which not $\Omega$ but $-\Omega$ is used. This results in
\begin{equation}
    \underline{d} = ( \cos{- \Omega t} \sin{i}, \cos{-\Omega t} \sin{i},\cos{i}).
\end{equation}

%\newpage

%--------FIGURE--------
%----------------------

%--------FIGURE--------
%----------------------

\section{Visibilities}
\label{app:vis}
%--------FIGURE--------
\begin{figure*}
\centering
		\includegraphics[width=13cm]{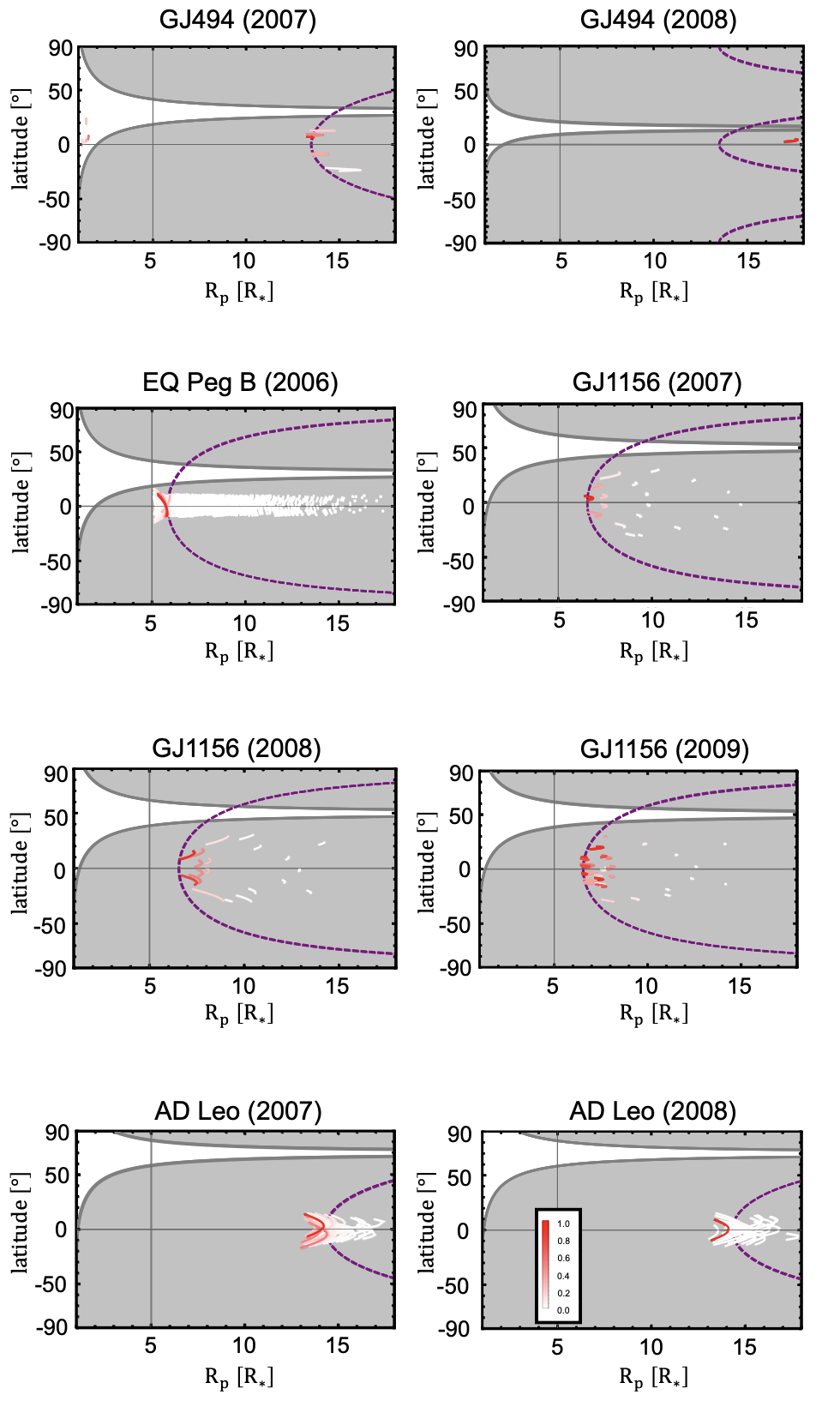}
	\caption{Visibility plots for stars in our sample, showing latitude against distance from the centre of the star ($R_p$). Any object lying in the white region would cross the stellar disk whilst the grey shaded regions are locations that could never be visible from Earth. The prominence formation sites are plotted and colour coded by mass that could be supported (scaled to the largest prominence mass of the map).}
    \label{fig:visibilities1}
\end{figure*}
%----------------------

%--------FIGURE--------
\begin{figure*}
\centering
		\includegraphics[width=12cm]{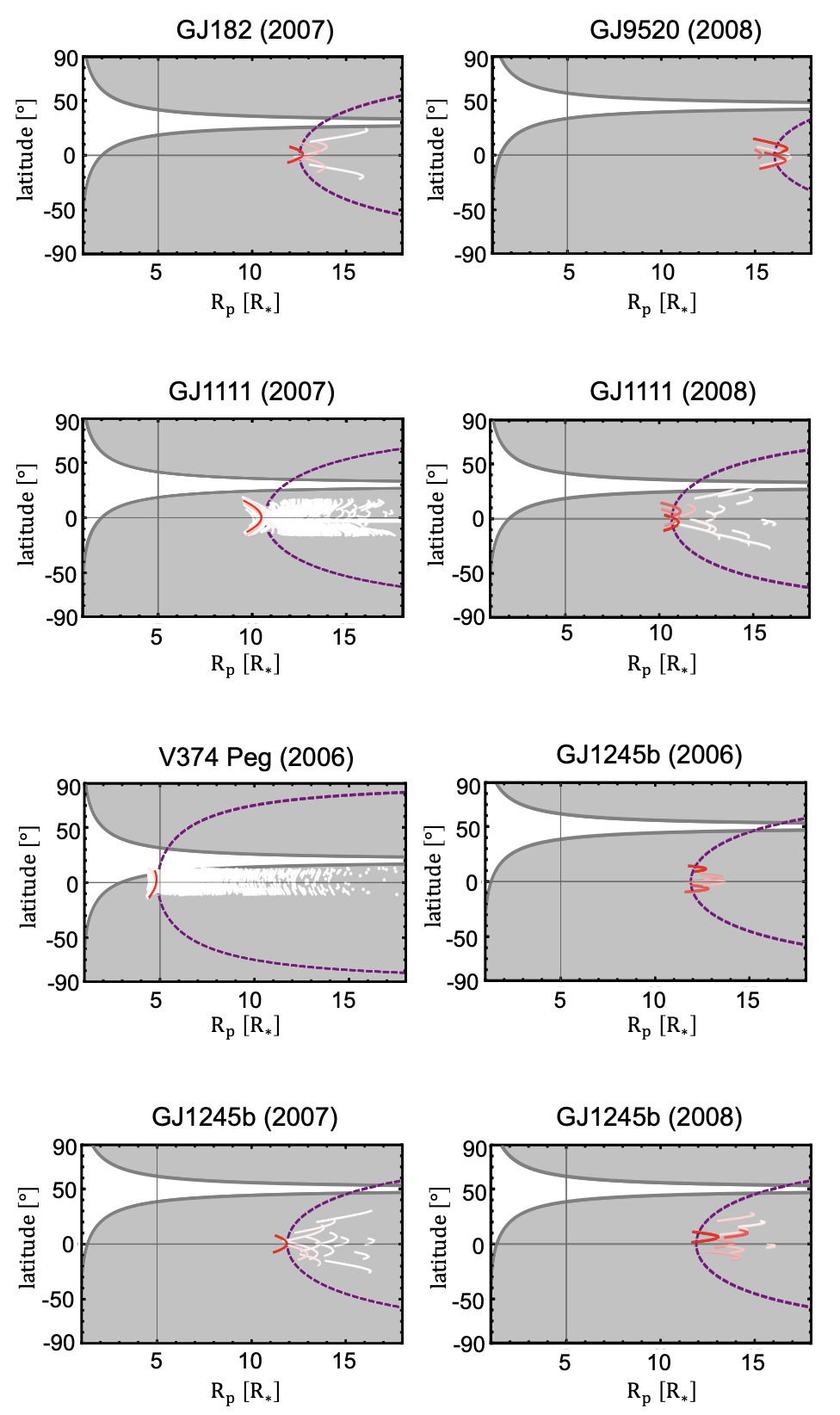}
	\caption{Visibility plots for the remaining stellar maps in our sample.}
    \label{fig:visibilities2}
\end{figure*}
%----------------------
%%%%%%%%%%%%%%%%%%%%%%%%%%%%%%%%%%%%%%%%%%%%%%%%%%

% Don't change these lines
\bsp	% typesetting comment
\label{lastpage}
\end{document}